\documentclass[11pt]{article}

\usepackage{caption}
\usepackage{epsfig}
\usepackage{color}
\usepackage{psfrag}
\usepackage{verbatim}
\usepackage{amsopn}
\usepackage{appendix}
\usepackage{dsfont,amsmath,amssymb,color,units,pstricks}
\usepackage{amsfonts}
\usepackage{graphicx}
\usepackage{natbib}
\usepackage{array}
\usepackage{lscape}
\usepackage{fullpage}
\usepackage{cancel}
\usepackage{url}
\usepackage{subcaption}
\allowdisplaybreaks

\newcommand{\cov}{{\rm cov}}
\newcommand{\corr}{{\rm corr}}

\newcommand{\Ex}{{\mathbb E}}

\newcommand\be{\begin{equation}}
\newcommand\bea{\begin{eqnarray} \nonumber }
\newcommand\ee{\end{equation}}
\newcommand\eea{\end{eqnarray}}

\providecommand{\keywords}[1]{\textbf{\textit{Keywords---}} #1}
\providecommand{\JEL}[1]{\textbf{\textit{JEL classification---}} #1}

\graphicspath{{./figures_fundamental/}}

\begin{document}

\unitlength = 1mm
\title{A continuous and efficient fundamental price \\ on the discrete order book grid}

\author{Julius Bonart$^{1}$\footnote{Corresponding author: j.bonart@ucl.ac.uk}, Fabrizio Lillo$^{2}$}

\maketitle

\noindent\small{$1$: Department of Computer Science, University College London and CFM-Imperial Institute of Quantitative Finance, Imperial College London}\\
\noindent\small{$2$: Scuola Normale Superiore, Piazza dei Cavalieri 7, Pisa, Italy} \\ 

\begin{abstract}
  This paper develops a model of liquidity provision in financial markets by adapting the Madhavan, Richardson, and Roomans (1997) price formation model to realistic order books with quote discretization and liquidity rebates. We postulate that liquidity providers observe a fundamental price which is continuous, efficient, and can assume values outside the interval spanned by the best quotes. We confirm the predictions of our price formation model with extensive empirical tests on large high-frequency datasets of 100 liquid Nasdaq stocks. Finally we use the model to propose an estimator of the fundamental price based on the rebate adjusted volume imbalance at the best quotes and we empirically show that it outperforms other simpler estimators.
\end{abstract}


\JEL{G10}

\newpage

\section{Introduction}

We say that a market price is \emph{efficient} if it unambiguously reflects all available information, is arbitrage-free, and unpredictable. But different types of microstructural frictions prevent the observed price to freely reflect the efficient price. Among these, the price discretization implemented in most modern markets plays a major role. As a result there is no reason to expect that either best quote, or the mid-price (their average), coincide with an efficient fundamental market price. 

Efficient price dynamics are somewhat paradoxical because many other financial time series display long memory: A striking example is the auto-correlation of the transaction signs \citep{Hasbrouck:correlation1987, Biais:ParisBourse1995, gefen, FarmerLillo04}. A central endeavour in market microstructure is to reconcile the predictability of trade signs with price efficiency, i.e. to uncover the adequate price formation process. Following \citet{glosten}, \citet{MRR} (henceforth MRR) developed a simple theory of price formation: Prices are impacted proportionally to the innovation in the transaction history (order flow). This alone removes all price predictibility when transaction signs are correlated. Included in their theory is \citet{glosten}'s assumption that market makers avoid ex-post regrets by setting a finite spread. Accordingly, the theory yields relationships between price impact, trade sign correlations, and the bid-ask spread. 

Because MRR's model disregards the role of price discretization it is not a priori clear that MRR's predicted relationships between market variables hold also for large tick stocks\footnote{In a large tick stock the ratio between tick size and price is relatively large and the spread is almost always equal to one tick.}. Hitherto, only some of these predictions have been empirically assessed in the literature. We perform here the first systematic test of MRR's general relationships and explicitly study how the tick size affects their performance. Although MRR's model assumed quotes and prices to be continuous we find that some of the predicted relationships between market variables are surprisingly accurate also when the tick is large and discretization plays, a priori, an important role.

More specifically, in this paper we shall argue that MRR's original idea can be adapted to large tick stocks by postulating the existence of an underlying efficient continuous market price. From this assumption we derive price formation equations for large tick stocks which reproduce, on average, the classical MRR relations. Our framework also predicts that price discretization becomes in fact important for second order price statistics (i.e. covariances and correlations). Thus, our model of liquidity provision explains how relationships between quantities depending linearly on the price hold regardless of the price discretization, whereas the tick size heavily influences quantities depending quadratically on the price.

In our model, limit order queues deplete when the fundamental price moves outside a certain interval around the mid price. We find that the length of this characteristic interval is typically larger than one tick and the mid-price dynamics are characterised by an interesting ``stickiness''. We show that this behaviour can be entirely explained by the existing rebate structure offered by the exchange to liquidity providers.

Our framework provides a straightforward way to reconcile the assumption of an efficient fundamental price with the inherent properties of financial markets. Whereas many empirical studies have focused on market dynamics on ultra high frequency time scales, little attention has been paid to the spatial dimension of price dynamics in modern financial markets. Because most liquidity provision in financial markets is nowadays channelled through high frequency market makers, important aspects of the market quality are naturally related to the properties of price changes on the smallest temporal and spatial scales. Our work develops an important step towards a unified theory of price formation and liquidity provision in small and large tick order books. 
Our point of view differs significantly from the perspective taken in some of the previous literature on the subject, too. We believe that our empirical observations suggest that price changes are mostly gouverned by strategic considerations of liquidity providers who agree on a hidden fundamental price. For example, we observe that a queue depletion at the best entails with a high probability a permanent price change. Order book models with purely stochastic dynamics, so called ``zero intelligence models'', are not capable of reproducing this simple fact, unless they benefit from additional assumptions. 

In large tick stocks the volumes at the best quotes contain a significant amount of information about the direction of the market~\citep{Gould:2015imbalance}. The second part of our paper hence proceeds by defining an approximate fundamental price in large tick stocks by using the \emph{squares} of the available volumes at the best bid and ask. This proxy can take continuous values within and beyond the region spanned by the bid and ask. We argue that this quantity performs better, as a proxy of the fundamental price, than generalized linearly volume weighted prices previously discussed in the literature. We assess the performance of our proxy by showing that (i) it incorporates to a large extent the information conveyed by the state of the order book, (ii) it behaves very much as the fundamental price, in that it follows approximately the MRR price formation rule. 
Having this proxy at hand allows us to reach further than much of the previous price formation literature, in that we can filter out exogeneous public information shocks from the fundamental price dynamics, and study their effects on the \emph{future} order flow. An interesting empirical finding of the second part of this paper is that the sign of the next trade is positively correlated with the exogeneous information shocks.

Tick sizes are not expected to decrease in the near future, since regulation authorities express their explicit desire to ``prevent a race to the bottom''\citep{gov_science}. European regulation agencies currently aim to establish a pan-European tick size regime ``to prevent the use of the tick size as a competition tool''\citep{MiFIDII} within the MiFID\footnote{The Markets in Financial Instruments Directive is the EU legislation regulating firms providing services to clients linked to 'financial instruments' (shares, bonds, units in collective investment schemes, and derivatives), and the venues where those instruments are traded.}. The introduction of MiFID II requires that standardized tick size bands, which depend on the valuation of the regulated security, must be adopted across Europe. In the United States, the SEC announced in a press release on May 6 2015 that it ``approved a proposal [...] for a two-year pilot program that would widen the minimum quoting and trading increments...'', i.e. the tick size, for stocks of some smaller companies~\citep{SECrelease}. This reflects a general mood inclined towards freezing the tick sizes or even increasing them. Such opinion of regulation agencies is supported by some academic research, although no clear consensus seems to have emerged \citep{Biais:survey2005}.
To summarize, we expect that the problem studied in this paper will have a relatively high longeveity and continue to be a challenge to regulators and market participants over a reasonably long term future. 

This paper is structured as follows:
In Sec.~\ref{sec:review} we disucss the relevant literature. Sec.~\ref{sec:LOB} reviews the functioning of limit order books. Sec.~\ref{sec:data} presents the empirical datasets used in this study. We analyse the empirical implications of the MRR price formation model in Sec.~\ref{sec:history}. Sec.~\ref{sec:mid} discusses the efficiency of the mid price in large tick LOBs and the role of liquidity rebates. Sec.~\ref{sec:theory} presents our generalization of the MRR model to large tick LOBs. We analyse first order and second order price statistics in small and large tick LOBs. A proxy of the fundamental price is developped in Sec.~\ref{sec:proxy}. Finally, in Sec.~\ref{sec:conclusion} we draw some conclusions.

\section{Literature review}
\label{sec:review}

The MRR model proposed a price formation process which accomodates the strong serial correlations of the trade signs \citep{Hasbrouck:correlation1987, Biais:ParisBourse1995, gefen, FarmerLillo04, toth15, Gould15a} with the very weak auto-correlation of price returns.
The MRR price formation model heavily relies on the theory of market making developed in \citet{glosten}. In recent years, \citet{glosten}'s work was significantly refined, for example in \citet{Wyart08,Bouchaud:2009digest}. However, despite the considerable recent effort in this field, no convincing theory for large tick stocks has emerged so far. History dependent impact was discussed again in~\citet{FarmerLillo04, farmer:2006fixed_or_temporary}. The expressions ``asymmetric/adaptive liquidity provision'' emerged to account for the impact asymmetry as a function of the order sign predictor~\citep{taranto}. MRR's transaction dependent price formation model was merged with this framework in~\citet{Bouchaud:2009digest}, albeit with some reluctance to use the concept of a ``fundamental price''.

In the econometrics literature the discrepancy between the efficient price and the quoted price is called market microstructural noise. The past literature mostly addressed the problem of filtering the market microstructural noise to uncover underlying quantities, such as feedback reactions between market participants~\citep{Bacry:2013}, fundamental price jumps~\citep{Lee:2015}, the volatility~\citep{Andersen:2010, Bandi:2008, Ghysels:2010,Jacod:2009}, and the influence of the tick size on the noise~\citep{Large:2010,Curato:2015}. \citet{Ball:truespreads2001} discussed the influence of the tick size on the adverse selection component of bid-ask spreads.

\citet{Rosenbaum:uncertainty, rosenbaum2} proposed an explicit method to reconstruct an efficient price from discrete high frequency data. Some aspects of our paper superficially resemble their work, in that we also propose a method to reconstruct a fundamental price from high frequency LOB data. However, in contrast to the method exposed in \citet{Rosenbaum:uncertainty} we shall achieve the reconstruction by using the best quotes and the displayed volumes in the LOB. \citet{Rosenbaum:uncertainty} use only the previous return and an ad-hoc quantity ``$\eta$'' to infer an efficient price. The parameter $\eta$ corresponds to the excess width of ``uncertainty regions'' where the efficient price is undefined. While this leads to mid price stickiness similar to what we observe in our model, we do not believe that the characteristic intervals around the mid-price reflect any uncertainty of liquidity providers about the fundamental price. Rather, we suggest that market making can be profitable even when the distance between the fundamental price and the mid price is larger than a half tick because of market making rebates offered by the exchanges; volumes are therefore not immediately cancelled when the fundamental price moves outside the interval between the bid and the ask, and the mid price becomes sticky. We follow here \citet{Harris:rebates} who discusses the consequences on market making and taking rebates/fees on the quote setting in limit order markets. Finally, \citet{Jaisson:fairprice} developed a model in which a fundamental price emerges as a consequence of a fair pricing condition for market makers. 

Alternatives to MRR's price formation theory emerged, as well. \citet{gefen} introduced the ``propagator model'' postulating a transient impact which was independent of the transaction history. Whereas \citet{Bouchaud:2009digest} showed that the transaction dependent propagator model was equivalent to the MRR theory, the multi event propagator model~\citep{eisler} truly differs from the MRR framework~\citep{taranto:2016}. The propagator model is an alternative to the MRR theory in that it completely lacks the notion of a fundamental price. 

Most classical price formation models, including the MRR framework and \citet{gefen}'s propagator model, have in common that they take the exogeneity of the order flow for granted. Our paper calculates explicitly the correlation between a public information shock and the \emph{future} trade sign in large tick order books. Therefore, our work contributes significantly to the understanding of the interplay between liquidity provision and consumption in financial markets. 
Superficially, this aspect of our study resembles \citet{Hasbrouck91}'s vector regression model of price returns and order flow and its recent generalization to large tick stocks ~\citep{taranto:2016b}. While we are able to infer directly the correlation between trade and public information shocks by using the proxy of the fundamental price, \citet{Hasbrouck91} performs an regression on price returns and order flow omitting the effect of public information shocks.

\section{Description of the order book mechanism}
\label{sec:LOB}

Most modern financial markets operate limit order books (LOBs), where financial institutions interact via the submission of orders. A buy order (sell order) is a commitment to buy (sell, respectively) a maximum quantity of the asset for a price no larger than its limit price (no smaller than its limit price, respectively). Whenever an institution submits a buy (sell) order, the LOB's trade-matching algorithms checks whether its order can be matched against previously submitted but still unmatched sell (buy) orders. In this case an immediate transactions occurs. If the order cannot be matched it remains active in the LOB until it is matched against a future incoming sell order or cancelled by its owner.
Orders that do not match upon arrival are called \emph{limit orders}. Orders that match upon arrival are called \emph{market orders}. 

For a given LOB, the \emph{bid price} $b_t$ is the highest price among active buy orders at time $t$. Similarly, the \emph{ask price} $a_t$ is the lowest price among active sell orders at time $t$. The bid and ask prices are collectively known as the \emph{best quotes}. Their difference $s_t=a_t-b_t$ is called the \emph{bid-ask spread,} and their mean $x_t=(a_t+b_t)/2$ is called the \emph{mid price}. 

\begin{figure}
\centering
\includegraphics[width=0.5\textwidth]{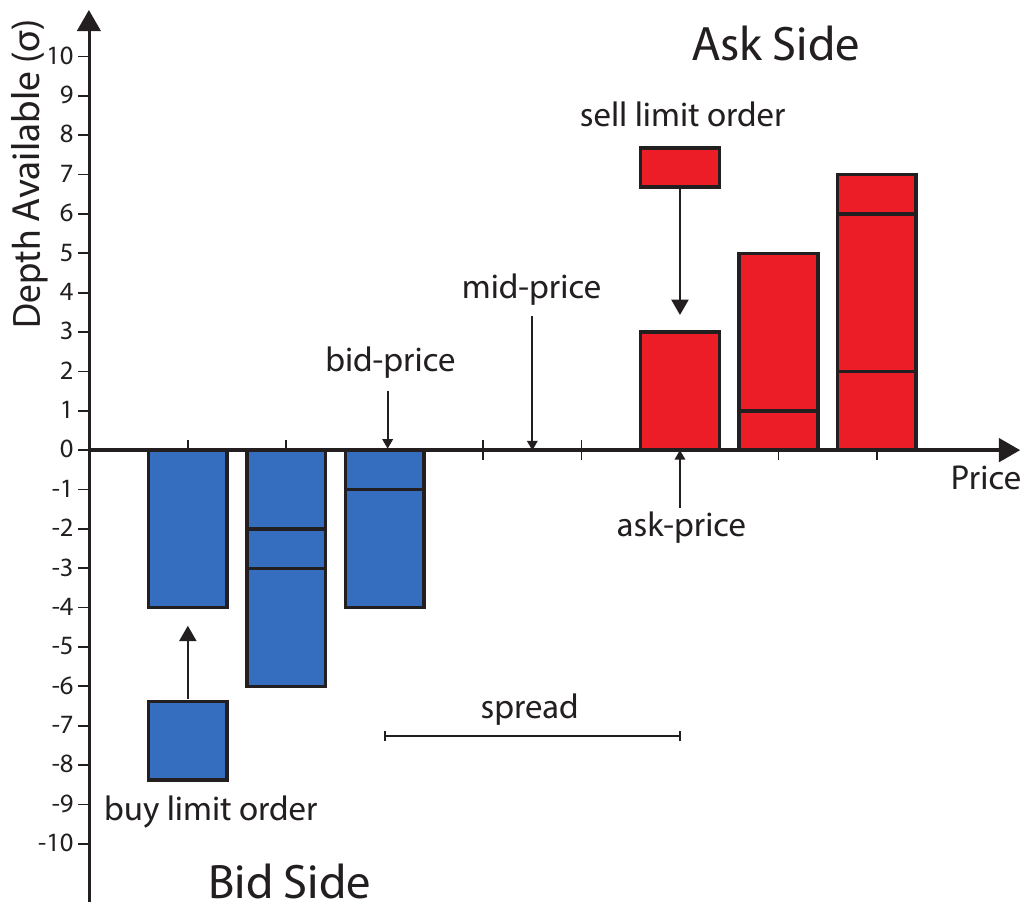}
\caption{Schematic of an LOB. The horizontal lines within the blocks at each price level denote the different active orders at each price.}
\label{fig:schematic}
\end{figure}

LOBs enforce a minimum price increment which is called the \emph{tick size}. Hence, institutions must choose prices of their orders which are integer multiples of the \emph{tick size} specified by the platform. Because LOBs implement a tick size $\tau>0$, it is common for several different limit orders to reside at the same price at a given time. A trade-matching algorithms therefore needs a secondary rule to decide which limit order among limit orders at the same price is executed first. To determine the queueing priority for orders at a given price, most exchanges implement a \emph{price--time} priority rule. That is, for buy (respectively, sell) limit orders, priority is given to the limit orders with the highest (respectively, lowest) price, and ties are broken by selecting the limit order with the earliest submission time. LOBs with price--time priority resembles in many respects a queueing system, where limit orders have a rank and wait for execution.
Figure \ref{fig:schematic} shows a schematic of an LOB at some instant in time, illustrating the definitions in this section.

The rules that govern order matching also dictate how prices evolve through time. Prices changes can occur for the following reason: The remaining volume at the best queue can be cancelled by its owner, or it can be executed by an incoming market order. Finally, if $s_t>\tau$ a buy or a sell limit order can be submitted inside the spread. In either case, the mid price changes by a certain mutiple of $\tau/2$. Both liquidity providers and liquidity consumers\footnote{A trader is a \emph{liquidity provider} if he/she submits limit orders. A trader is a \emph{liquidity consumer} if he/she submits market orders.} can therefore change the price in a LOB.

When the tick size is small, the effects of the discretization are weak and can often be neglected. When the tick size is large, the effects of the discretization are important and constrain the price dynamics. For example, the spread of large tick LOBs is usually locked and equal to the allowed minimum of one tick, whereas the spread of small tick LOBs is usually a large multiple of $\tau$. Because price fluctuations are proportional to the price itself, not the bare tick size itself but the ratio $\tau/x_t$ determines whether an asset is a large or small tick. We call this ratio the \emph{relative tick size} $\tilde \tau$. 

Finally, market participants pay fees to the exchange when they trade in a LOB. Some exchanges (e.g. Nasdaq) also offer rebates to liquidity providers: Owners of limit orders receive a negative fee (i.e. credit) from the exchange upon the execution of their orders as a reward for liquidity provision.

\section{Data}
\label{sec:data}

In our empirical investigation we use a pool of 100 highly liquid stocks traded on Nasdaq during the whole year 2015. The data that we study originates from the LOBSTER database \citep{LOBSTER}, which lists all market order arrivals, limit order arrivals, and cancellations that occur on the Nasdaq platform during the normal trading hours of $09$:$30$ to $16$:$00$ on each trading day. Trading does not occur on weekends or public holidays, so we exclude these days from our analysis. We also exclude market activity during the first and last hour of each trading day, to remove any abnormal trading behaviour that can occur shortly after the opening auction or shortly before the closing auction.

On the Nasdaq platform, each stock is traded in a separate LOB with price--time priority, with a tick size of $\tau = \$0.01$. Although this tick size is the same for all stocks on the platform, the prices of different stocks vary across several orders of magnitude (from about $\$1$ to more than $\$1000$). Therefore, the \emph{relative tick size} similarly varies considerably across different stocks. The bid-ask spread of large tick LOBs is typically close to one tick (its minimum possible value). In this paper we define large tick stocks by the condition $\Ex [s_t] < 0.013\$ = 1.3\tau$, with $\Ex [s_t] $ the observed time averaged bid-ask spread. Conversely, the bid-ask spread of small tick LOBs is typically a large multiple of the tick. In this paper we define small tick stocks by the condition $\Ex [s_t]  > 0.04\$ = 4\tau$. All other stocks are called medium tick stocks.

Nasdaq imposes fees of approximately $0.003\$$ per share on liquidity takers and offers rebates of approximately $0.003\$$ per share to liquidity providers \citep{nasdaq:fees}. 

The LOBSTER data has many important benefits that make it particularly suitable for our study. First, the data is recorded directly at the Nasdaq servers. Therefore, we avoid the many difficulties associated with data sets that are recorded by third-party providers, such as misaligned time stamps or incorrectly ordered events. Second, each market order arrival listed in the data contains an explicit identifier for the limit order to which it matches. This enables us to perform one-to-one matching between market and limit orders, without the need to apply inference algorithms for this purpose, which can produce noisy and inaccurate results. Third, each limit order described in the data constitutes a firm commitment to trade. Therefore, our results reflect the market dynamics for real trading opportunities, not ``indicative'' declarations of possible intent.
In table~\ref{tab} we provide summary statistics of each stock in the pool.

\section{Transaction history dependent price formation}
\label{sec:history}

\emph{Price formation} describes the process whereby information, liquidity, and the order flow impact the market price. The price impact mechanism cannot be trivial: It is a very well known empirical fact that the order flow is highly correlated~\citep{Hasbrouck:correlation1987, Biais:ParisBourse1995}\footnote{The main reason of the long memory in the market order flow is order splitting. Institutions tend to split large orders into small orders and execute them incrementally over long periods of time. Order flow correlations are detectable on time scales beyond hours and even days \citep{FarmerLillo04, gefen, toth15, lillo, Bouchaud:2009digest}.} whereas the resulting mid prices are (nearly) uncorrelated.

Madvahan et al. \citet{MRR} developed a simple structural model of price formation. They postulated the existence of a fundamental price\footnote{More precisely MRR define $p_t$ as the ``post trade expected value of the stock conditioned upon public information and the trade information variable".} $p_t$ of the stock. $t$ denotes the \emph{transaction time}, i.e. $p_t$ is the fundamental price immediately before the $t$-th trade. If a transaction at time $t$ is buyer initiated, it is assigned the indicator $\epsilon_t = 1$, while if it is seller-initiated it is assigned the indicator $\epsilon_t = -1$. MRR further assumed that the revision in beliefs about $p_t$ is positively correlated with the innovation in the order flow. Formally, this can be written as
\begin{equation}\label{eq:pt}
  p_{t+1} - p_t = G[\epsilon_t - \hat\epsilon_t] + W_t\;,
\end{equation}
where
\begin{equation}
  \hat\epsilon_t = \Ex_{t-1}[\epsilon_t]
\end{equation}
is the expected transaction sign at $t$ given the public information up to $t$. Public information enters the market through the past trading history $\{\epsilon_{t-1}, \epsilon_{t-2},\cdots\}$ and the shocks $\{W_{t-1}, W_{t-2}, \cdots\}$ which describe public informaton derived from external news\footnote{MRR originally assumed that $\hat\epsilon_t$ only depended on $\epsilon_{t-1}$. We consider a more general case in this paper as in \citep{FarmerLillo04,taranto}.}. We assume that $W_t$ is a white noise with zero mean which is uncorrelated with the past transaction history\footnote{The future trade signs can be correlated with $W_t$ because market participants might adapt their trading strategy to the observed price dynamics. We revert back to this problem in Sec.~\ref{sec:secondorder}.}. Private  information enters Eq.~(\ref{eq:pt}) through the term $G[\epsilon_t - \hat\epsilon_t]$. Because the information on the past order flow is public, the information content of a trade can only depend on its unexpected part $\epsilon_t - \hat\epsilon_t$. 
By construction, this price formation mechanism ensures price efficiency, $\Ex_{t-1}[p_{t+1}]=p_t$, regardless the correlations of $\{\epsilon_t\}$. 

Market makers share a common belief about $p_t$ and set ask and bid quotes. Following \citet{glosten} they seek to avoid ex-post regrets. In a competitive market, the bid and ask are therefore
\begin{align}
  a_t &= p_t + G[1 - \hat\epsilon_t] \;,\label{eq:at}\\
  b_t &= p_t + G[-1- \hat\epsilon_t] \;,\label{eq:bt}
\end{align}
which ensure a zero average gain for the market maker\footnote{Zero average gain is the idealized state of perfect competition. In reality, market makers need to set slightly larger bid-ask spread to achieve a minimum profit, albeit this profit is arguably very low.}. Accordingly, the LOB's mid-price and spread are
\begin{align}
  x_t &= \frac{a_t+b_t}{2} = p_t - G\hat\epsilon_t \;,\label{eq:xt}\\
  s_t &= a_t - b_t = 2G \;.\label{eq:st}
\end{align}
Because market makers need to anticipate the impact of the future market order, the mid price at $t$ is not equal to $p_t$. More specifically, its evolution equation reads
\begin{equation}\label{eq:xt2}
  x_{t+1} - x_t = G[\epsilon_t - \hat \epsilon_{t+1}] + W_t\;.
\end{equation}
MMR studied this evolution equation and compared the implied spread to actually observed spreads. In their conclusion, they pointed out that price discreteness was a serious limitation to the applicability of their model. Surprisingly, this has been hitherto taken for granted in the market microstructure literature. Moreover, confusion arised later whether Eq.~(\ref{eq:pt}) described a fundamental price or the observed mid-price \citep{Bouchaud:2009digest}, since the consequences are manifestly different. Hence, the empirical implications of the evolution equation~(\ref{eq:xt2}) have not been fully analysed in the literature. Our goal is therefore to transform Eq.~(\ref{eq:xt2}) into easily observable quantities. 

\begin{figure}
  \centering
  \includegraphics[scale=0.5]{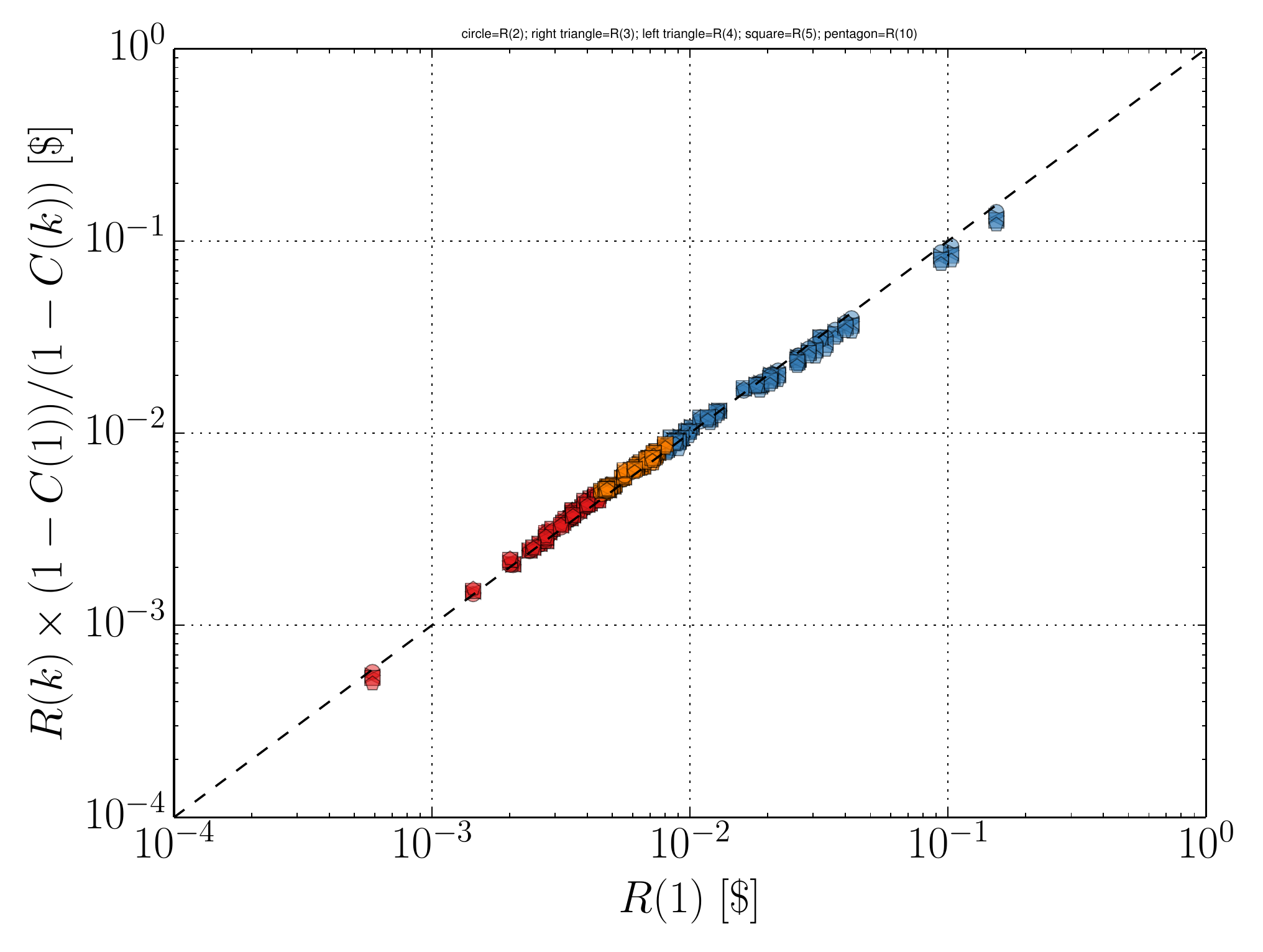}
  \caption{Emprirical test of the prediction~(\ref{eq:Rt_R1}) of the MRR model on a pool of $100$ large, medium and small tick stocks. Each cloud of points, consisting of a circle (rescaled $R(2)$), a right triangle (rescaled $R(3)$), a left triangle (rescaled $R(4)$), a square (rescaled $R(5)$), and a pentagon (rescaled $R(10)$) corresponds to a single stock, averaged over the year 2015. Red clouds depict large tick stocks (which we define by $\Ex [s_t]<0.013\$$), yellow clouds correspond to medium tick stocks (which we define by $0.013\$ < \Ex [s_t]  < 0.04\$$, and blue clouds correspond to small tick stocks (which we define by $\Ex [s_t]  > 0.04\$$). The black dotted line ($y=x$) is a guide to the eye.}
  \label{Rt_versus_R1}
\end{figure}

We first note that the response function of a transaction, defined as  \citep{gefen} 
\begin{equation}
  R(\ell) = \Ex[\epsilon_t\cdot(x_{t+\ell} - x_t)]\;,
\end{equation}
is related to the correlations in the order flow according to
\begin{align}
  R(\ell)&= G\Ex[1 - \epsilon_t\hat\epsilon_{t+\ell}] = G - G\Ex\left[\epsilon_t \cdot \Ex_t(\hat\epsilon_{t+\ell}-\epsilon_{t+\ell})\right] - G\Ex[\epsilon_t\epsilon_{t+\ell}]\nonumber\\
  &= G[1 - C(\ell)]\;.\label{eq:RC}
\end{align}
We exploited the fact that by definition $\Ex_t[\hat\epsilon_{t+\ell} - \epsilon_{t+\ell}] = 0$ and defined the auto-correlation function of the trade signs
\begin{equation}
  C(\ell) = \Ex[\epsilon_{t+\ell}\epsilon_t]\;.
\end{equation}
Both $C(\ell)$ and $R(\ell)$ are easily measurable on empirical data. A testable parameter-free prediction of the MRR model\footnote{The following equation does not appear in the original MRR paper, but it is mentioned in \citet{Wyart08}.} is that
\begin{equation}\label{eq:Rt_R1}
  R(1) = R(\ell) \frac{1 - C(1)}{1 - C(\ell)} \;.
\end{equation}
The results of this empirical test are displayed in Fig.~\ref{Rt_versus_R1}. Our dataset reproduces the relationship~(\ref{eq:Rt_R1}) accurately over three orders of magnitude. This is highly surprising, as it contains both small and large tick stocks, and price discretization implies that the equations for $x_t$ and $s_t$ (Eqs.~(\ref{eq:xt})-(\ref{eq:st})) cannot be exact. Yet, Eq.~(\ref{eq:Rt_R1}) holds also for large tick stocks. Even the stock with the largest relative tick in our dataset, Sirius XM Holdings, satisfies Eq.~(\ref{eq:Rt_R1}) approximately, see table~\ref{tab:siri}.
\begin{table}
  \centering
  \begin{tabular}{c|c|c}
    & SIRI & MSFT \\
    \hline
    $R(1)$  & 0.000588 \$ & 0.00277 \$ \\
    $R(2)\frac{1-C(1)}{1-C(2)}$  & 0.000571 \$ & 0.00287 \$ \\
    $R(3)\frac{1-C(1)}{1-C(3)}$  & 0.000548 \$ & 0.00294 \$ \\
    $R(4)\frac{1-C(1)}{1-C(4)}$  & 0.000539 \$ & 0.00299 \$  \\
    $R(5)\frac{1-C(1)}{1-C(5)}$  & 0.000531 \$ & 0.00302 \$  \\
    $R(10)\frac{1-C(1)}{1-C(10)}$  & 0.000497 \$ & 0.00305 \$ \\
    $R(20)\frac{1-C(1)}{1-C(20)}$ & 0.000489 \$ & 0.00299 \$  \\
    \hline
  \end{tabular}
  \caption{Empirical test of the relation (\ref{eq:Rt_R1}) of the MRR model for Sirius XM Holdings, the stock with the largest relative tick in our dataset, and Microsoft Corp., the most liquid large tick stock in our dataset.}
  \label{tab:siri}
\end{table} 

In the following we show with minimal assumptions why the testable MRR predictions~(\ref{eq:Rt_R1}) are correct also on large ticks. Our framework will show that while Eq.~(\ref{eq:xt2}) is certainly incorrect as it stands, a very similar version remains true \emph{on average}.

\section{Price formation in large tick stocks}
\label{sec:theory}

We postulate the existence of a fundamental price which takes continuous values and satisfies the MRR equation~(\ref{eq:pt}). When the tick size is large, the best bid and ask are separated by one tick: $a_t-b_t = \tau$. Eqs.~(\ref{eq:bt}) and~(\ref{eq:at}) are now incorrect because they do not account for price discretization.

In large tick LOBs the mid price changes when limit order queues at the best deplete. We assume that liquidity providers own volume at the best quotes as long as it is marginally profitable, i.e. as long as its expected execution price is equal to the fundamental price immediately after the transaction, adjusted for the liquidity rebate.
A limit order at the best remains profitable as long as
\begin{equation}\label{eq:ptregion}
  p_t \in (b_t - r - G[-1-\hat\epsilon_t], a_t + r - G[1-\hat\epsilon_t])\;,
\end{equation}
where $r > 0$ is the market making rebate per share offered by the exchange, $p_t + G[-1-\hat\epsilon_t]$ is the expected fundamental price after the execution of the buy limit order at $b_t$, and $p_t + G[1-\hat\epsilon_t]$ is the expected fundamental price after the execution of the sell limit order at $a_t$. 
Nasdaq offers rebates for liquidity providers of approximately $0.003\$$ per share which ensures that a limit order at the best can be profitable to its owner even when the fundamental price is outside the interval $(b_t, a_t)$. We confirm this prediction empirically in Sec.~\ref{sec:mid} by showing that both the impact of an extreme volume imbalance and the impact of a queue depletion on the mid price is approximately equal to the rebate adjusted half spread.

Price discretization prevents MRR's dynamic equation for $x_t$, Eq.~(\ref{eq:xt2}), to be correct, but we can show that it is still true \emph{on average}. We assume that the distribution of $p_t$ within the characteristic interval is symmetric with respect to its center. In this case the expected fundamental price, given the mid price, is according to Eq.~(\ref{eq:ptregion}), equal to the average of the edges of the characteristic interval around $x_t$:
\begin{equation}
  \Ex[p_t|x_t] = x_t + G\hat\epsilon_t\;.
\end{equation}
We can apply the conditional expectation to the above relation, given the past until $t'\le t$, and average over $x_t$. This is particularly useful, since Eq.~(\ref{eq:pt}) implies that $\Ex_{t'}[p_{t+1} - p_t] = G \Ex_{t'}[\epsilon_t - \hat\epsilon_t]$ for $t'\le t$. This allows us to express the return of the mid price in terms of known quantities:
\begin{equation*}
  G \Ex_{t'}[\epsilon_t - \hat\epsilon_t] = \Ex_{t'}[x_{t+1} - x_t] + G \Ex_{t'}[\hat\epsilon_{t+1} - \hat\epsilon_t]\;,
\end{equation*}
or simply
\begin{equation*}
  \Ex_{t'}[x_{t+1} - x_t] = G \Ex_{t'}[\epsilon_{t} - \hat\epsilon_{t+1}]\;.
\end{equation*}
Note that $t'\le t$ is arbitrary.
Therefore, we can sum up this equation to obtain
\begin{equation}\label{eq:xt3}
  \Ex_t[x_{t+\ell} - x_{t}] = G\Ex_t[\epsilon_{t} - \hat\epsilon_{t+1+\ell}]\;.
\end{equation}
We can now proceed by calculating the response function of a transaction, 
\begin{equation}
  R(\ell) = \Ex[\epsilon_t\cdot(x_{t+\ell} - x_t)] = G\Ex\left(\epsilon_t\cdot\Ex_t[\epsilon_t - \hat\epsilon_{t+\ell+1}]\right) = G\Ex\left(\Ex_t[1 - \epsilon_t\epsilon_{t+\ell+1}]\right) = G[1 - C(\ell)] \;.
\end{equation}
This is exactly the MRR relation~(\ref{eq:RC}), which we have so successfully tested on empirical data. Because MRR assumed a continuous price scale, it was not clear why its predictions turned out to be true for large tick LOBs as well. Whereas price discretization implies that the MRR equation~(\ref{eq:xt2}) is incorrect as it stands, our Eq.~(\ref{eq:xt3}) demontrates that it nevertheless holds true \emph{on average}: Relationships between quantities which depend \emph{linearly} on the mid price are thus not affected by the tick size.

In Sec.~\ref{sec:secondorder} however we introduce the auto-covariance function of mid price returns. Because the auto-covariance depends \emph{quadratically} on the mid price, the mathematical expectation in Eq.~(\ref{eq:xt3}) leads to results which are very different from what the orginal MRR Eq.~(\ref{eq:xt2}) predicts.

\subsection{Volume imbalance at the best and mid price dynamics}
\label{sec:mid}

If we accept that a fundamental price should reflect publicly available information, then it is easy to show \emph{empirically} that the mid price cannot be the efficient fundamental price. Define the volume imbalance
\begin{equation}
  \iota(t) = \frac{V_b(t) - V_a(t)}{V_b(t) + V_a(t)}\;,
\end{equation}
where $V_a(t)$ and $V_b(t)$ denote the available volumes at the ask and bid price. This volume imbalance has attracted some attention in the recent literature~\citep{Gould:2015imbalance,Donnelly, CarteaJaimungal15,queue-reactive}. The quantity $\iota$ can be interpreted as the pressure that liquidity providers put on the price, in that when $\iota>0$, more limit orders have been submitted and not canceled at the buy side, and prices are expected to go up. When $\iota<0$, more limit orders have been submitted at the sell side and prices are expected to go down. 
We can confirm this intuition quantitatively by considering the price impact of an imbalance, namely
\begin{equation}
  R(\ell|\iota) = \Ex[ x_{t+\ell} - x_t|\iota(t) = \iota]\;,
\end{equation}
which is the expected price change after $\ell$ transactions given that an imbalance $\iota$ has been observed at $t$. 

\begin{figure}
  \centering
  \includegraphics[scale=0.45]{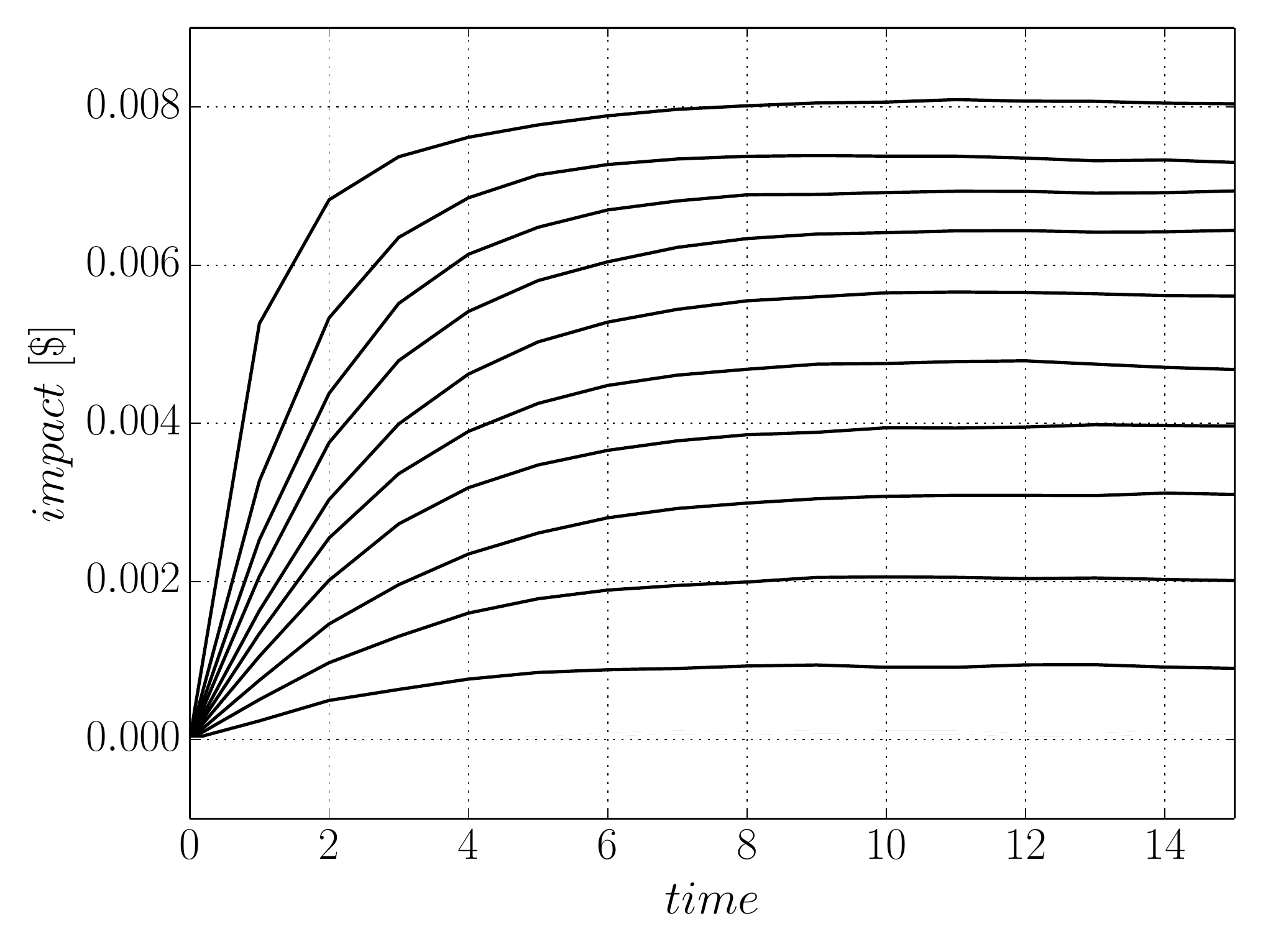}
  \caption{The symmetrized impact of an imbalance on $x_t$ for Microsoft and $|\iota|$ contained in $10$ equidistant bins; $|\iota| \in [0.1\times k, \; 0.1\times (k+1))$, for $k = 0,1,\cdots,9$. $k$ increases from the bottom to the top line.}
  \label{fig:imb}
\end{figure}

If $x_t$ was efficient all price predictors constructed with public data would have zero impact: In particular, the response function conditioned on the queue imbalance $\iota$  would vanish, $R(\ell|\iota) = 0$. This is however in stark contrast with empirical observations: Fig.~\ref{fig:imb} shows $R(\ell|\iota)$ as a function of $\ell$ for different intervals of values of $\iota$. As soon as $\iota\ne 0$ the subsequent expected price change is non zero. Interestingly, for large $|\iota|$ the expected absolute price change exceeds a half tick. We also find empirically that $R(\infty|\iota=1) = -R(\infty|\iota=-1)$ is approximately equal to the average absolute impact of a queue depletion on the midprice, denoted by $r_\infty(t)$, in agreement with intuition, see table~\ref{tab:imp_of_dep}.

The existence of (at least) one price predictor $\iota$ which can have an impact larger than $\tau/2$ suggests that the true underlying price can be outside the region spanned by $b_t$ and $a_t$ because $\Ex[p_{t+\ell}|\iota(t)] = p_t$ irrespective of $\iota(t)$ by construction and $\Ex[x_\infty|p_t] = p_t$.
Therefore, the absolute permanent impact of a queue depletion at $t$ is given by 
\begin{equation} 
  r_\infty(t) = |p_{t^+} - x_t| = |p_t + G(\epsilon_t - \hat \epsilon_t) - x_t|  = |\pm\left(\frac{\tau}{2} + r\right) - G(\epsilon_t-\hat\epsilon_t) + G(\epsilon_t-\hat\epsilon_t)| = \frac{\tau}{2} + r\;,
\end{equation} 
with $p_{t^+}$ the post-trade fundamental price and $x_t$ the pre-trade mid price. Note that $r_\infty(t)$ can be larger than a half tick. Nasdaq offers rebates for liquidity providers of approximately $0.003\$$ per share. The average permanent impact of a queue depletion is thus
\begin{equation}
  r_\infty = \frac{\tau}{2} + r = 0.008\$ \;.
\end{equation}
In Table~\ref{tab:imp_of_dep} we compare the impact of queue depletion to the half tick adjusted by the market making rebate.

In conclusion, since $r_\infty$ is in general larger than $\tau/2$, the fundamental price can lie outside the interval $(b_t,a_t)$. The mid-price is therefore ``sticky'': It remains constant as long as $p_t$ is within a certain interval around $x_t$. Because these characteristic intervals reside on the price scale with periodicity $\tau$, but overlap for different $x_t$, the mid price $x_t$ can take two different values for the same $p_t$, depending on the past dynamics of $p_t$ (see \citet{Rosenbaum:uncertainty} for a similar observation in the context of high-frequency econometrics).

\begin{table}
  \centering
  \begin{small}
  \begin{tabular}{c|c|c}
    \hline
    ticker & permanent impact of & permanent impact of\\
           & depletion $[0.01\$]$ & large imbalance $[0.01\$]$ \\
    \hline
    AMAT & 0.81 (0.015) & 0.76 (0.011)\\
    ATVI & 0.78 (0.015) & 0.83 (0.017)\\
    CA & 0.76 (0.013) &  0.85 (0.019) \\
    CMCSA & 0.78 (0.009) &  0.88 (0.011)\\
    CSCO & 0.81 (0.010) & 0.76 (0.008) \\
    CSX & 0.76 (0.011) &  0.88 (0.015) \\
    DISCA & 0.71 (0.011) & 0.94 (0.024)\\
    EBAY & 0.80 (0.016) & 0.89 (0.015)\\
    FOX & 0.77 (0.012) &0.88 (0.018) \\
    GE & 0.72 (0.013) &0.67 (0.011) \\
    INTC &0.80 (0.010) & 0.76 (0.008) \\
    JPM & 0.67 (0.007)  &0.86 (0.013) \\
    MAT & 0.82 (0.015)  &0.87 (0.018) \\
    MDLZ & 0.79 (0.012) & 0.85 (0.014) \\
    MSFT & 0.83 (0.010) &0.79 (0.007)\\
    MU & 0.76 (0.010) &0.74 (0.011) \\
    NVDA & 0.80 (0.024) &0.83 (0.014) \\
    ORCL & 0.86 (0.025) &0.86 (0.014) \\
    QCOM & 0.72 (0.008) &0.90 (0.014) \\
    SIRI &0.65 (0.047) & 0.66 (0.017) \\
    SYMC & 0.82 (0.015) &0.85 (0.015)\\
    TXN &0.80 (0.017) & 0.92 (0.016) \\
    VOD &0.78 (0.014) & 0.83 (0.016) \\
    YHOO &0.80 (0.009) & 0.83 (0.010) \\

    \hline
  \end{tabular}
  \end{small}
  \caption{Comparison between the permanent price impact of a queue depletion and the permanent price impact of a large imbalance ($|\iota| > 0.9$) for all large tick stocks in our dataset, defined by the condition $\Ex[s_t] < 0.013\$$. Both values are close to the rebate adjusted half tick of $0.5 + 0.3 = 0.8$ dollar cents. Standard deviations are shown in parantheses.}
  \label{tab:imp_of_dep}
\end{table}

\subsection{Covariance of returns}
\label{sec:secondorder}

We have shown that the mid price is not efficient in large tick LOBs because it does not incorporate the information conveyed by the volume imbalance. When the tick size is small, the volume imbalance loses its predictive power \citep{Gould:2015imbalance}. A direct way of assessing the efficiency of the mid price in small tick LOBs is to consider the empirical covariance function of mid price returns
\begin{equation}
  \cov_x(\ell) = \Ex[(x_{t+1+\ell} - x_{t+\ell})(x_{t+1} - x_t)] \;.
\end{equation}
The mid price is not efficient, neither in small tick nor in large tick LOBs, because we observe that the covariance function of lag one is in most cases negative, i.e. the mid price mean reverts after a non-zero return.

In this section we show that MRR's relation (\ref{eq:xt2}) allows us to forecast the covariance of mid price returns for small tick stocks. But the mathematical expectation in the corresponding equation~(\ref{eq:xt3}) of our model for large ticks prevents us from forecasting the covariance of mid price returns in large tick stocks and we shall explicitly observe this empirically.

The continuous price MRR's model can be easily adapted to finite liquidity rebates. When market makers earn rebates they avoid ex-post regrets by setting quotes according to $a_t = p_t + G[1-\hat\epsilon_t] - r$ and $b_t = p_t + [-1-\hat\epsilon_t] + r$. Yet, the mid price $x_t = \frac{1}{2}[b_t+a_t]$ is unaffected by $r$ and we can use Eq.~(\ref{eq:xt2}) to calculate
\begin{equation*}
  \cov_x(\ell) = G^2\left(C(\ell) - C(\ell+1) - \Ex[\epsilon_{t+\ell}\hat\epsilon_{t+1}] + \Ex[\epsilon_{t+\ell+1}\hat\epsilon_{t+1}]\right)
  + G \Ex[\epsilon_{t+\ell}W_t] - G\Ex[\epsilon_{t+\ell+1}W_t]
\end{equation*}
It is important to realize that we do not assume that the future order flow is uncorrelated with $W_t$\footnote{We find empirically that $\cov_x(1)<0$ for many stocks. MRR assumed in their original paper that the trade signs were Markovian and exogenous. In this case one has $\hat\epsilon_{t+1}=C(1)\epsilon_t$ which implies $\cov_x(1) = G^2C(1)[1-C(1)]^2>0$, in contradiction with our empirical findings. On the other hand, if we assume that the trade sign process is exogenous, but allows for a ``perfect'' predictor, i.e. $\hat\epsilon_{t+1}=\epsilon_{t+1}$, we find that $\cov_x(1) = G^2[2C(\ell) - C(\ell+1) - C(\ell-1)]$ which is \emph{negative} when $C(\ell)$ is concave. Therefore, even a purely exogenous order flow is compatible with a negative auto-covariance of returns.}. In fact, it is intuitively clear that market participants react to past returns. Order flow and price dynamics are therefore connected through a feedback loop,as shown empirically in \citep{Hasbrouck91} and more recently with a different approach in \citep{taranto:2016, taranto:2016b}.

Despite this difficulty it is possible to obtain a non-parametric prediction of the covariance of mid price returns. We define the trade sign shifted response function as
\begin{equation}\label{eq:Rell1}
  R_k(\ell) = \Ex[\epsilon_{t+k}(x_{t+\ell} - x_t)] \;.
\end{equation} 
$R_k(\ell)$\footnote{Note that $R_0(\ell) = R(\ell)$ is the standard reponse function of a transaction. $R_\ell(\ell) = -R(-\ell)$ is minus the response function with negative lag considered in \citet{taranto:2016}.} is easily measurable on empirical data. Surprisingly, by using the MRR equation~(\ref{eq:xt2}) it is possible to express the covariance of mid price returns entirely in terms of trade sign shifted response functions:
\begin{equation}\label{eq:test}
  \cov_x(\ell) = G\left[R_\ell(1) - R_{\ell+1}(1)\right] = R(\infty)\left[R_\ell(1) - R_{\ell+1}(1)\right]\;.
\end{equation}
This relation is a subtle test of Eq.~(\ref{eq:xt2}) in that it is valid irrespective the correlations of the price dynamics (i.e. the noise $W_t$) with the future order flow.

\begin{figure}
  \centering
  \includegraphics[scale=0.397]{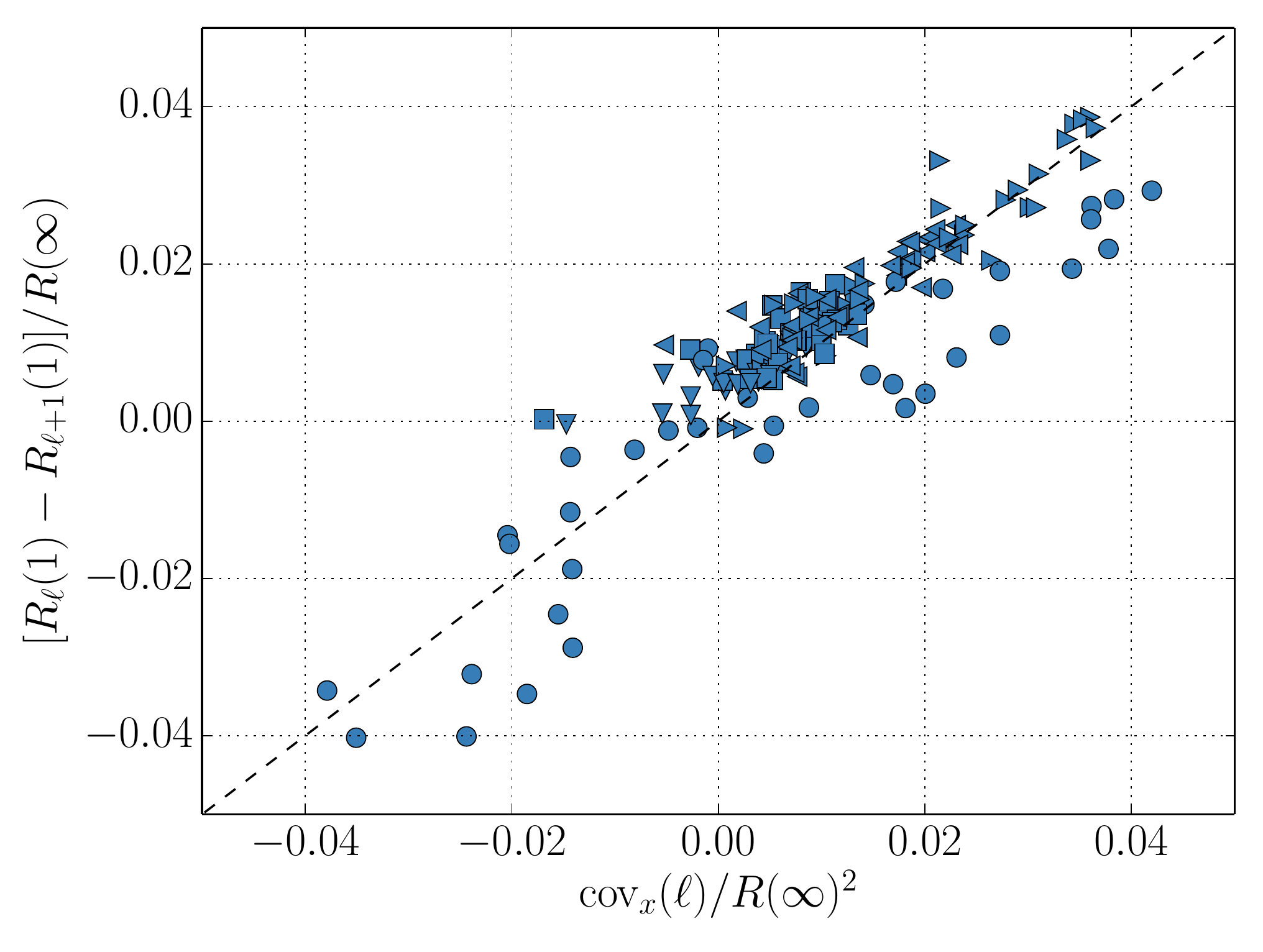}
  \includegraphics[scale=0.397]{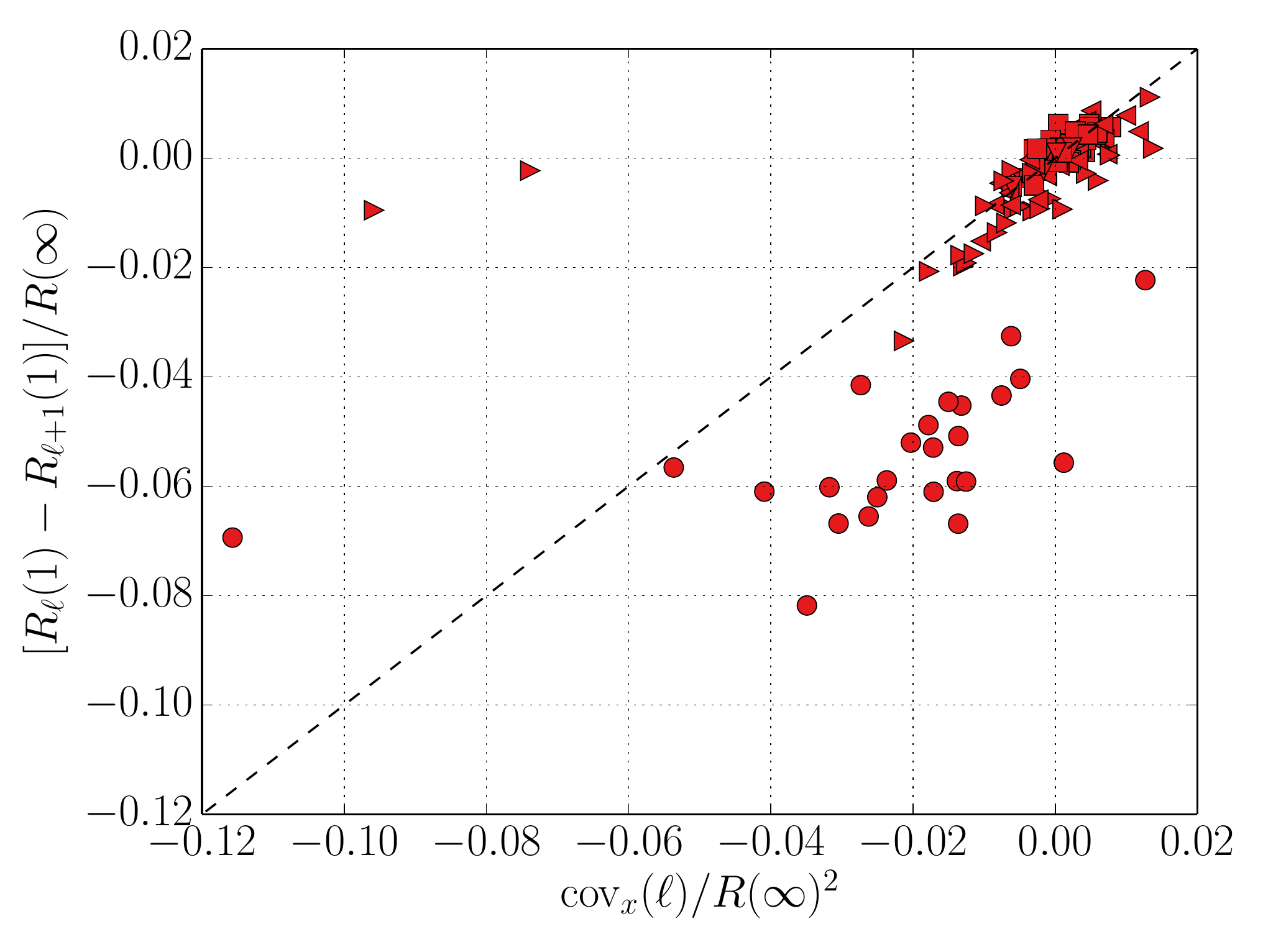}
  \caption{The auto-covariance of mid price returns, rescaled by the permanent market order impact $R(\infty)$, for (left figure) all small tick (defined by a spread $\Ex [s_t] > 0.04\$$) and for (right figure) all large tick stocks (defined by a spread $\Ex [s_t]  < 0.013\$$) in our dataset. Each stock's $\cov_x(1)$ is marked as a circle, the $\cov_x(2)$ are marked as right triangles, the $\cov_x(3)$ are marked as left triangles, the $\cov_x(4)$ are marked as squares, and each stock's $\cov_x(5)$ is marked as a down triangle.}
  \label{fig:cov1}
\end{figure}

When the relative tick size is large, our framework predicts that Eq.~(\ref{eq:xt2}) holds only on average. Because the covariance function is a second order statistics, we therefore do not expect Eq.~(\ref{eq:test}) to be true for large tick stocks.
In Fig.~\ref{fig:cov1} we test Eq.~(\ref{eq:test}) on our empirical dataset. We observe that Eq.~(\ref{eq:test}) performs relatively well for small tick stocks and relatively badly for large tick stocks, as we have predicted. 

In summary then, we conclude that, first, the linear dependence between the price evolution and private and public information in the classical MRR model describes the price dynamics in small tick LOBs accurately, in that it performs very well with respect to linear and moderately well\footnote{Note that in reality the transaction \emph{sizes} are not constant. Because large transactions have in general a larger impact than small transactions, the prefactor $G$ depends on the size and fluctuates in time. Whereas this does not play a role in linear price statisics, the time correlations of $G$ do change the second order statistics. Our model neglects this effect.} with respect to quadratic price statistics. Second, this statement is true regardless the correlation between the information shocks and the \emph{future} order flow. Finally, our generalized model for large tick LOBs correctly predicts that the price discreteness is irrelevant regarding linear price statistics, but has a significant influence on quadratic price statistics.

\section{Estimators of the fundamental price for large tick stocks}
\label{sec:proxy}

The question remains if and how the fundamental price can be reconstructed from publicly available LOB data. This section introduces a proxy $\hat p_t$ of the fundamental price in \emph{large tick} LOBs, which we construct by using the \emph{squared volumes} at the best: 
\begin{equation}\label{eq:hatpt}
  \hat p_t = \frac{V^2_a(t)(b_t - r) + V^2_b(t)(a_t + r)}{V^2_a(t) + V^2_b(t)}\;.
\end{equation}
$\hat p_t$ has the following desirable properties: It is easy to calculate and continuous. Its values can lie outside the interval $(b_t,a_t)$ as required in our framework. Finally, in a balanced LOB, i.e. when $V_a(t) = V_b(t)$, it coincides with the mid price.

Below we present several additional criteria which motivate our choice of using Eq.~(\ref{eq:hatpt}). To substantiate our choice of $\hat p_t$, we compare its performance to the alternative proxy
\begin{equation}\label{eq:hatptilde}
  \hat p'_t = \frac{V_a(t)(b_t - r) + V_b(t)(a_t + r)}{V_a(t) + V_b(t)}\;,
\end{equation}
which is the simplest linear generalization of the standard volume weighted price
\begin{equation}\label{eq:hatptilde2}
  \hat p_t'' = \frac{V_a(t)b_t + V_b(t)a_t}{V_a(t)+V_b(t)}\;.
\end{equation}
Note that $\hat p_t''$ cannot lie outside the interval $(b_t,a_t)$. By showing that $\hat p_t$ performs better than $\hat p''_t$ we can thus substantiate one of the key aspects of our model.

We choose to measure the performance of $\hat p_t$, $\hat p_t'$, and $\hat p_t''$ by considering: (i) the impact of an imbalance on the price, (ii) the unconditional response function, and (iii) the autocorrelation of returns, which we also investigate by looking at the signature plot of volatility. Our empirical results can be summarized as follows.

\begin{figure}
  \centering
  \includegraphics[scale=0.45]{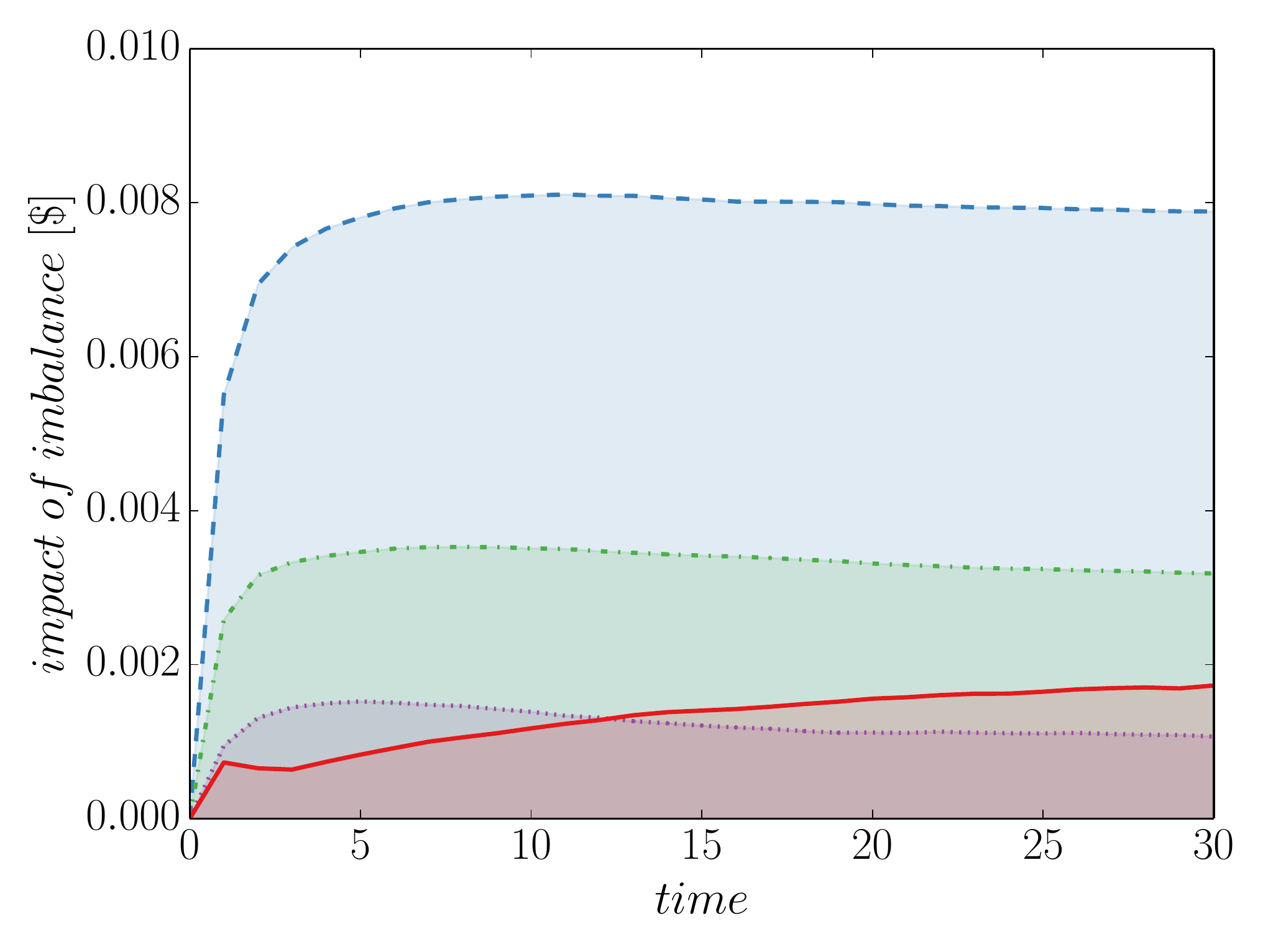}
  \caption{The symmetrized impact of imbalances $|\iota| \in [0,1]$ on the proxy $\hat p_t$ of the fundamental price defined in Eq.~(\ref{eq:hatpt}) (region below red solid), on the proxy $\hat p'_t$ defined in Eq.~(\ref{eq:hatptilde}) (region below violet dotted), on the proxy $\hat p''_t$ defined in Eq.~(\ref{eq:hatptilde2}) (region below green dash-dotted), and on the mid price $x_t$ (region below blue dashed).}
  \label{fig:imp_of_fund}
\end{figure}

First, $\hat p_t$ has incorporated to a large extent the information conveyed by the volume imbalance between the ask and bid. To demonstrate this, we calculate the impact of an imbalance on $\hat p_t$ for Microsoft, the most liquid large tick stock in our dataset. We use Nasdaq's liquidity rebate of approximately $r = 0.3\tau$. The result is depicted in Fig.~\ref{fig:imp_of_fund}. We observe that the impact of imbalances on $\hat p_t$ is only $\approx 20\%$ as large as the impact on the mid price $x_t$. We interpret the missing impact as the information which is already included in $\hat p_t$. Thus, $\hat p_t$ captures approximately $80\%$ of the information conveyed by the volume imbalance. We also calculate the impact of an imbalance on the alternative proxies $\hat p_t'$ and $\hat p_t''$. While $\hat p_t'$ also captures approximately $80\%$ of the information of the imbalance, the proxy $\hat p_t''$ performs much worse. We interpret this fact as an indication that proxies of the fundamental price which are adjusted for liquidity rebates (i.e. which can lie outside the region spanned by the bid and ask) are significantly more efficient with respect to $\iota$. 

Second, $\hat p_t$ behaves approximately as the efficient fundamental price in the MRR framework, in that it approximately satisfies Eq.~(\ref{eq:pt}). Define the lagged response of a transaction on $\hat p_t$,
\begin{equation*}
  R^{(\hat p)}(\ell) = \Ex[\epsilon_t\cdot (\hat p_{t+\ell} - \hat p_t)]\;.
\end{equation*}
 If $\hat p_t$ was exactly equal to the fundamental price, $R^{(\hat p)}(\ell)$ would coincide with $R^{(\hat p)}(1)$ for all $\ell \ge 1$. 

\begin{figure}
  \centering
  \includegraphics[scale=0.45]{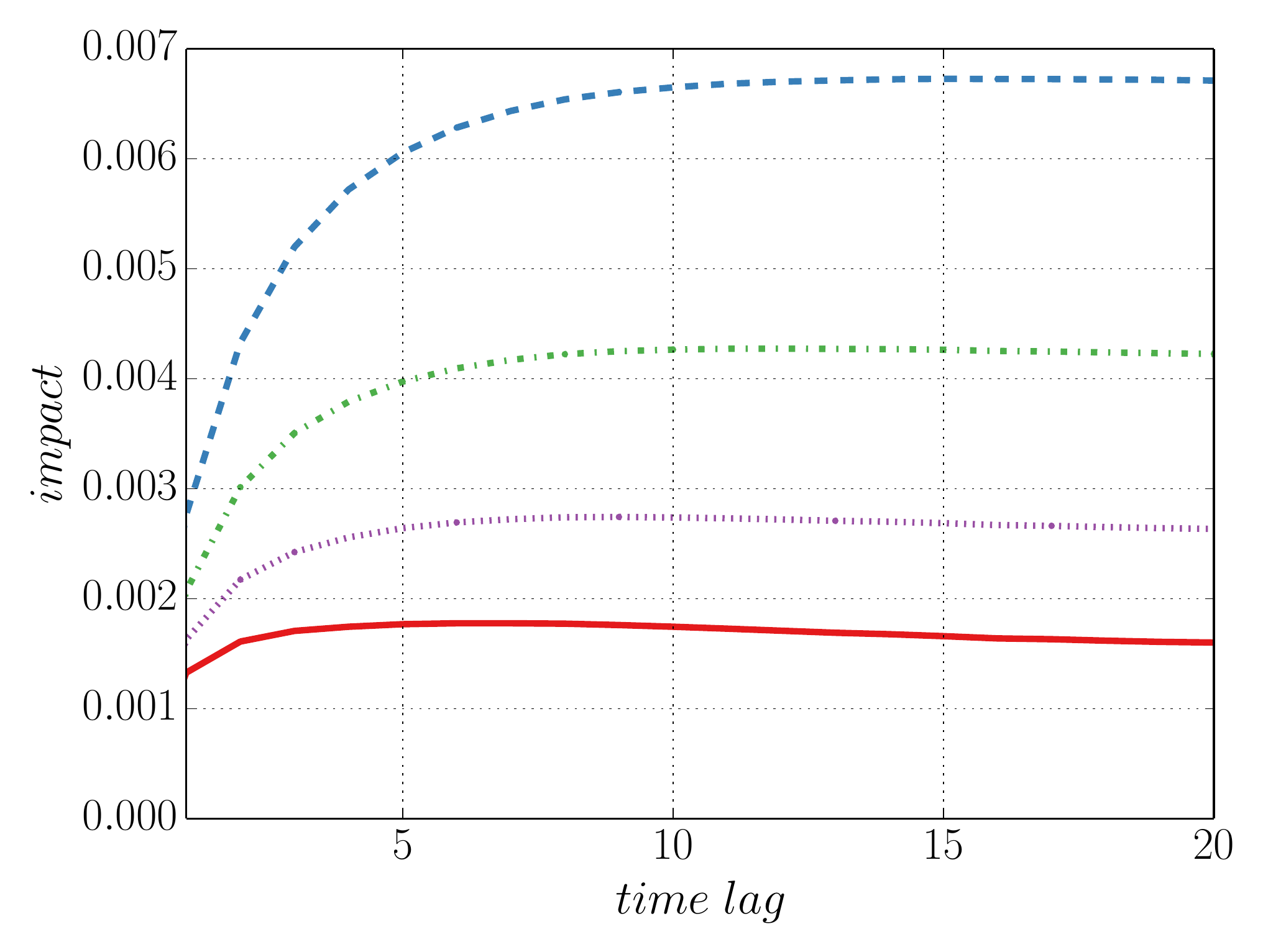}
  \caption{The response function of a trade on the proxy of the fundamental price Eq.~(\ref{eq:hatpt}), $R^{(\hat p)}(\ell) = \Ex[\epsilon_t\cdot (\hat p_{t+\ell} - \hat p_t)]$, for Microsoft (red solid) compared to the response function $R(\ell)$ calculated with the mid price (blue dashed), the response function $R^{(\hat p')}$ calculated with the alternative proxy Eq.~(\ref{eq:hatptilde}) (violet dotted), and the response function $R^{(\hat p'')}$ calculated with the alternative proxy Eq.~(\ref{eq:hatptilde2}) (green dash-dotted).}
  \label{fig:R_function_on_fundamental}
\end{figure}

In Fig.~\ref{fig:R_function_on_fundamental} we plot $R^{(\hat p)}(\ell)$, $R^{(\hat p')}(\ell)$, $R^{(\hat p'')}(\ell)$, and $R(\ell)$. Whereas the standard permanent impact, $R(\infty)$, differs by a factor of $2.5$ from $R(1)$, we observe that the difference between $R^{(\hat p)}(\infty)$ and $R^{(\hat p)}(1)$ is reduced to $\approx 10\%$. Moreover, the permanent impact on $\hat p_t$ is much weaker than the permanent impact on $x_t$. We suggest that this difference is observed because $\hat p_t$ incorporates a large part of the past and future order flow correlations. In fact, if $\hat p$ is the fundamental price, Eq.~(\ref{eq:pt}) predicts 
\begin{equation*}
  R^{(\hat p)}(\infty) = R^{(\hat p)}(1) = \Ex[\epsilon_t\cdot(p_\infty - p_t)] = G(1 - \Ex[\epsilon_t\cdot\hat\epsilon_t])\;,
\end{equation*}
which is, depending on the predictor $\hat\epsilon_t$, significantly smaller than $R(\infty) = G$.

\begin{figure}
  \centering
  \includegraphics[scale=0.45]{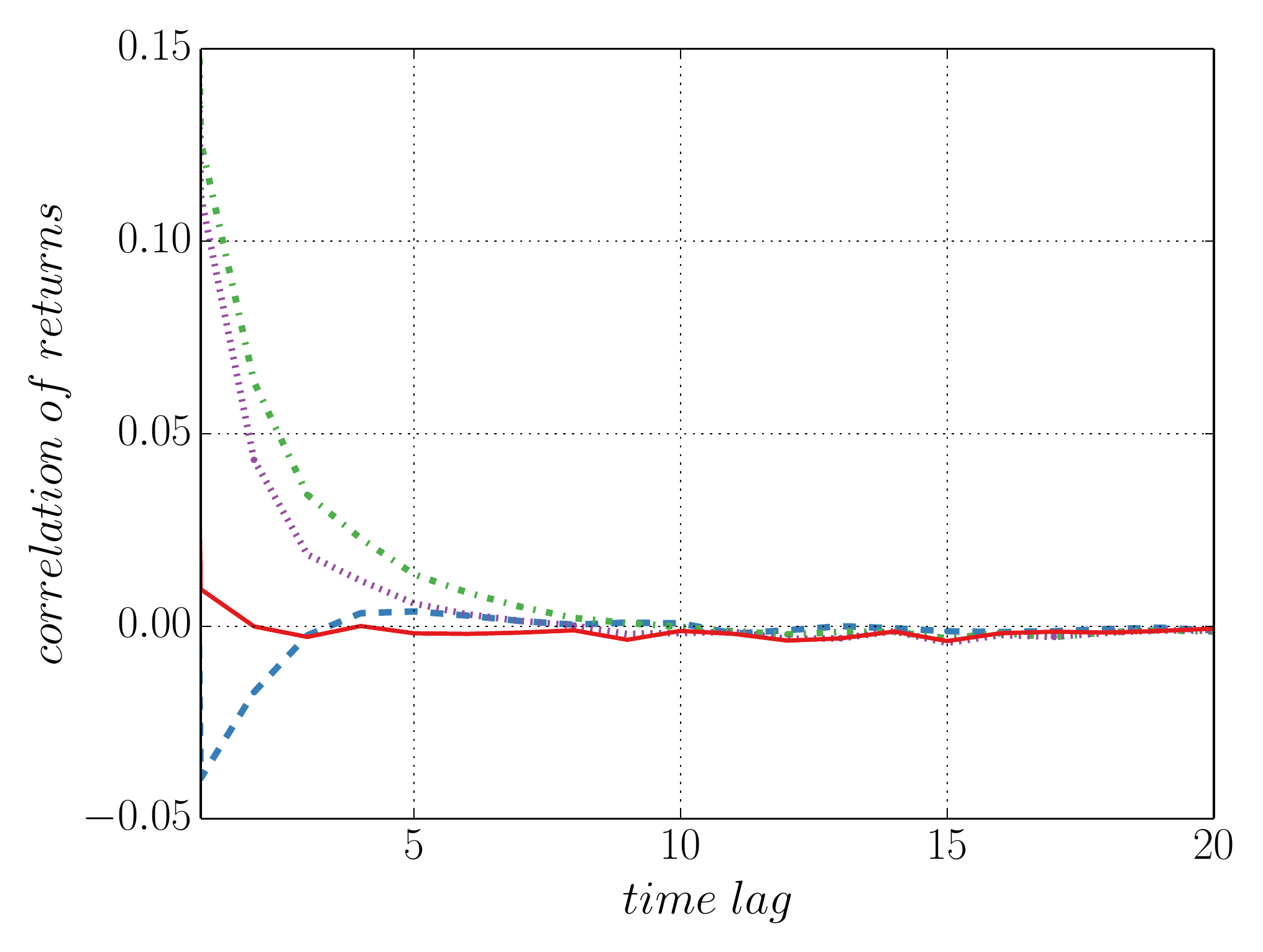}
  \caption{The correlation function of returns for Microsoft, calculated with the proxy $\hat p_t$ of the fundamental price defined in Eq.~(\ref{eq:hatpt}) (red solid), the alternative proxy $\hat p_t'$ defined in Eq.~(\ref{eq:hatptilde}) (violet dotted), the alternative proxy $\hat p_t''$ defined in Eq.~(\ref{eq:hatptilde2}) (green dash-dotted), and the mid price $x_t$ (blue dashed).}
  \label{fig:correl}
\end{figure}

Third, to test for statistical efficiency of $\hat p_t$,  we calculate the correlation function of returns separated by a time lag $\ell$:
\begin{equation}
  \corr_{\hat p}(\ell) = \frac{\Ex[(\hat p_{t+\ell+1} - \hat p_{t+\ell}) (\hat p_{t+1} - \hat p_t)]}{\Ex[(\hat p_{t+1}-\hat p_t)^2]} \;.
\end{equation}
We also consider the correlation functions $\corr_{\hat p'}(\ell)$, $\corr_{\hat p''}(\ell)$, and $\corr_{x}(\ell)$ which are defined accordingly. Fig.~\ref{fig:correl} shows that $C_{\hat p}(\ell)$ is approximately zero for all $\ell$. Therefore, $\hat p_t$ is virtually statistically unpredictable with linear methods, as required by Eq.~(\ref{eq:pt}). Fig.~\ref{fig:correl} shows also the auto-correlation functions of the returns of the alternative proxies $\hat p_t'$, $\hat p_t''$ and the mid price $x_t$. We observe that both $\hat p_t'$ and $\hat p_t''$ suffer from a relatively large positive auto-correlation reflecting our previous observation that the response functions $R^{(\hat p')}$ and $R^{(\hat p'')}$ continue to grow significantly after a transaction (see Fig.~\ref{fig:R_function_on_fundamental}). Table~\ref{tab:autocorrelation} summarizes our findings on the auto-correlations of $\hat p_t$, $\hat p_t'$, $\hat p_t''$ and $x_t$ for the large tick stocks. Section~\ref{sec:secondorder} discusses the causes of the negative auto-correlation function of the mid price returns.

\begin{table}
  \centering
  \begin{small}
  \begin{tabular}{c|c|c|c|c}
    ticker & $\corr_{\hat p}(1)$ & $\corr_{\hat p'}(1)$ & $\corr_{\hat p''}(1)$ & $\corr_{\hat x}(1)$\\
    \hline
    AMAT & 0.021 (0.003) & 0.114 (0.003) & 0.107 (0.003) & -0.030 (0.003) \\
    ATVI & 0.027 (0.003) &0.111 (0.004) &0.108 (0.003) & -0.026 (0.004)\\
    CA  & 0.041 (0.004) & 0.127 (0.004) &0.125 (0.004) & -0.014 (0.004) \\
    CMCSA & 0.033 (0.003) & 0.135 (0.003) &0.145 (0.003) & -0.021 (0.003) \\
    CSCO & -0.006 (0.002) & 0.086 (0.003) &0.115 (0.002) &  -0.004 (0.002)\\
    CSX & 0.040 (0.003) & 0.126 (0.003) &0.119 (0.003) &  -0.033 (0.003) \\
    DISCA & 0.064 (0.004) & 0.139 (0.004) &0.138 (0.004) & 0.014 (0.004) \\
    EBAY & 0.043 (0.003) & 0.138 (0.003) &0.143 (0.003) & 0.000 (0.003) \\
    FOX & 0.038 (0.003) & 0.126 (0.004) &0.128 (0.003) & -0.031 (0.004) \\
    GE & -0.045 (0.004) & 0.046 (0.005) &0.085 (0.004) & -0.058 (0.003) \\
    INTC &0.018 (0.002) & 0.112 (0.002) &0.119 (0.002) &-0.030 (0.002) \\
    JPM & 0.023 (0.003) & 0.124 (0.003) &0.140 (0.003) & -0.031 (0.003) \\
    MAT & 0.059 (0.004) & 0.143 (0.004) & 0.136 (0.004) & -0.011 (0.004)\\
    MDLZ & 0.033 (0.003) & 0.127 (0.004) &0.128 (0.003) & -0.021 (0.003)\\
    MSFT & 0.015 (0.002) & 0.117 (0.002) & 0.126 (0.002) & -0.040 (0.002) \\
    MU & 0.009 (0.004) & 0.100 (0.004) &0.114 (0.004) & -0.035 (0.003) \\
    NVDA &0.040 (0.003) & 0.130 (0.003) &0.126 (0.003) & -0.018 (0.003) \\
    ORCL &0.026 (0.003) & 0.129 (0.004) &0.137 (0.003) & -0.025 (0.003)\\
    QCOM &0.031 (0.003) & 0.127 (0.003) &0.134 (0.003) & -0.017 (0.003) \\
    SYMC &0.037 (0.003) & 0.128 (0.003) &0.119 (0.003) & -0.023 (0.003) \\
    SIRI &-0.03 (0.006) &0.013 (0.007) & 0.065 (0.006) & -0.015 (0.006) \\
    TXN &0.033 (0.003) & 0.119 (0.003) & 0.124 (0.003) & -0.018 (0.003) \\
    VOD &0.004 (0.003) & 0.087 (0.004) & 0.124 (0.003) & -0.038 (0.004) \\
    YHOO &0.022 (0.004) & 0.120 (0.004) & 0.133 (0.004) & -0.032 (0.004)\\
    \hline
  \end{tabular}
  \end{small}
  \caption{Auto-correlation functions at lag $1$ for the returns of the proxies $\hat p_t$, $\hat p_t'$ and $\hat p_t''$ of the fundamental price, and the auto-correlation at lag $1$ for the returns of the mid price for all large tick stocks (defined by $\Ex[s_t] <0.014\$$) in our dataset. Standard deviations are shown in parantheses.}
  \label{tab:autocorrelation}
\end{table}

\begin{figure}
  \centering
  \includegraphics[scale=0.5]{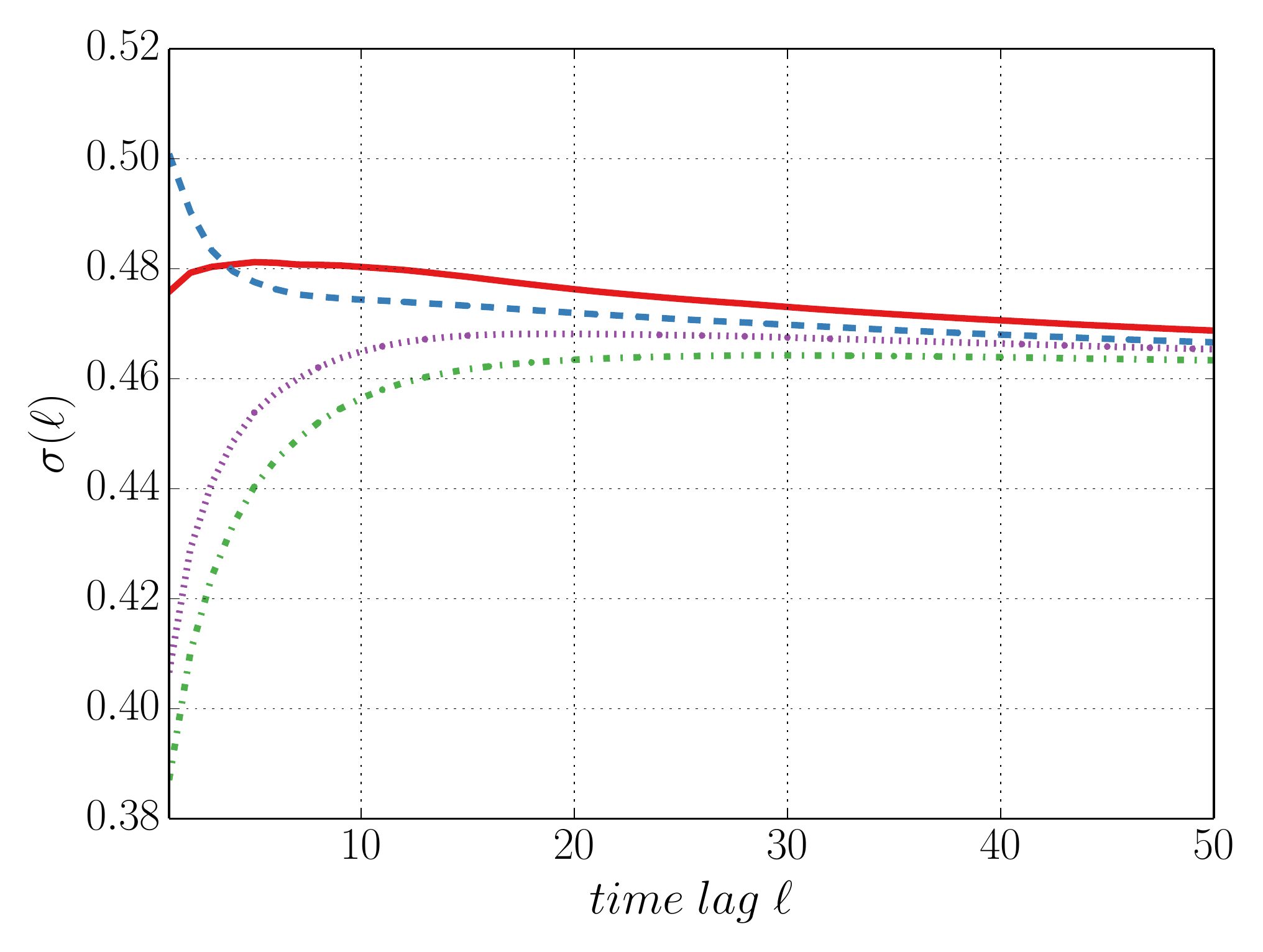}
  \caption{Signature plot of the proxies $\hat p_t$ (red solid), $\hat p_t'$ (violet dotted), $\hat p_t''$ (green dash-dotted), and the mid price $x_t$ (blue dashed) for Microsoft during the year 2015.}
  \label{fig:signature}
\end{figure}

A related method to test the  statistical efficiency of $\hat p_t$  is to investigate its signature plot which displays the time-normalized volatility defined as the mean square displacement of lag $\ell$ rescaled by $\ell$: 
\begin{equation*}
  \sigma_{\hat p}(\ell) = \sqrt{\frac{\Ex[(\hat p_{t+\ell} - \hat p_t)^2]}{\ell}} \;.
\end{equation*}
The time-normalized volatility of a martingale is independent of $\ell$, whereas an increasing time-normalized volatility indicates a positive auto-correlation of returns (trend following) and a decreasing time-normalized volatility indicates a negative auto-correlation of returns (mean-reversion). We also calculate the equivalent time-normalized volatilities for the other proxies $\hat p_t'$, $\hat p_t''$, as well as for $x_t$. Fig.~\ref{fig:signature} shows the signature plot of $\hat p_t$, $\hat p'_t$, $\hat p_t''$, and $x_t$. We observe again that $\hat p_t$ performs better than the alternative proxies in that its time-normalized volatility depends much less on the time lag.

Finally, we remark that $\hat p_t$ coincides, by construction, with the average true fundamental price in our model when a queue depletion is imminent, i.e. when either $V_b(t)$ or $V_a(t)$ are very close to zero. While we stress that the goal of this paper is not to seek for the best proxy of the fundamental price, we believe that the above analysis demonstrates that $\hat p_t$, as it is defined in Eq.~(\ref{eq:hatpt}), can serve as a  good approximation to the true fundamental price.

Having a proxy of the fundamental price at hand is useful for several reasons. First, it is a natural measure of the future direction of the market, in that the future average mid price is expected to increase if $x_t<\hat p_t$ and expected to decrease if $x_t>\hat p_t$. Thus, while the mid price does not coincide with the fundamental price due to market frictions (here the tick size), it nevertheless converges on average to the latter in the future. Deviations between the mid price and the fundamental price are thus transient, an intuition that we share with \citet{Hasbrouck91}. 

Second, $\hat p_t$ allows us to study the covariance between the next trade sign $\epsilon_{t+1}$ and the public information $W_t$ at $t$. Because the information process is exogeneous and uncorrelated from the past order flow, this covariance captures a genuine innovation of the beliefs shared amongst liquidity \emph{takers} due to the arrival of new information. The theory of price formation has been hitherto mostly concerned with the reaction of liquidity providers, i.e. \emph{quote setters}, to new information (for new developments see however \citep{taranto:2016, taranto:2016b}).
By using again the trade sign shifted response function, defined in Eq.~(\ref{eq:Rell1}), we have
\begin{equation}
  R^{(p)}(1) = G(1 - \Ex[\epsilon_t\hat\epsilon_t]) = R(1) + R_1(1) - \Ex[\epsilon_{t+1}W_t] \;,
\end{equation}
which allows us to measure the covariance between $W_t$ and $\epsilon_{t+1}$ by using the proxy $\hat p_t$ for $p_t$. Table~\ref{tab:news} contains our estimates of $\Ex[\epsilon_{t+1}W_t]$ for all large tick stocks in our dataset. We empirically observe that $\Ex [\epsilon_{t+1}W_t] > 0$. Thus, positive news trigger on average more orders to buy and negative news trigger on average more orders to sell, in agreement with intuition.
\begin{table}
  \centering
  \begin{small}
  \begin{tabular}{c|c}
    ticker & $\Ex \epsilon_{t+1}W_t \;[0.01\$]$ \\
    \hline
    AMAT  & 8.15 (0.45)\\
    ATVI  & 10.98 (0.60)\\
    CA  & 14.76 (0.66)\\
    CMCSA & 14.96 (0.52)\\
    CSCO & 8.42 (0.38)\\
    CSX & 14.75 (0.62)\\
    DISCA & 15.86 (0.70)\\
    EBAY & 12.02 (0.71)\\
    FITB & 11.29 (0.52)\\
    FOX & 18.17 (0.58)\\
    GE & 6.95 (0.55)\\
    INTC & 8.89 (0.38)\\
    JPM & 22.54 (0.50)\\
    MAT & 12.81 (0.54)\\
    MDLZ & 10.44 (0.63)\\
    MSFT & 9.25 (0.43)\\
    MU & 9.28 (0.51)\\
    NVDA & 11.17 (0.51)\\
    ORCL &13.17 (0.73)\\ 
    QCOM &12.98 (0.55)\\
    SIRI &1.66 (0.28)\\
    SYMC &10.72 (0.64)\\
    TXN &17.38 (0.56)\\
    VOD &25.55 (0.64)\\
    YHOO &14.91 (0.53)\\
  \end{tabular}
  \end{small}
  \caption{Implied covariance between public news $W_t$ and the sign of the next transaction $\epsilon_{t+1}$ for all large tick stocks (defined by their spread $\Ex [s_t]  < 0.013\$$) in out dataset. Standard deviations are shown in parantheses.}
  \label{tab:news}
\end{table}

\section{Conclusions}
\label{sec:conclusion}

While price discreteness is an inherent property of financial markets, the interplay between liquidity provision, spread dynamics, and information asymmetry has hitherto been often analysed under the assumption of a continuous price scale. 

We believe that this assumption is unsustainable in the light of the modern functioning of trading platforms: 
First, liquidity is nowadays mostly provided by high frequency market makers who seek to make a profit from tiny inefficiencies. Assuming that these inefficiencies are necessarily of the order of the LOB price resolution would be a gross mistake. Our paper shows precisely that changes and mismatches in the consensus price can be much smaller than the tick. 
Second, assuming continuous prices excludes the possibility of analysing the modern market design and its influences on the trading environment. The advantages and disadvantages of certain order priority rules, the influence of latency on the investor's welfare, the profitability of hidden liquidity, are some of the regulatory issues which can be fully addressed only by considering finite tick sizes.
 
This paper has overcome these limitations by developping a consistent model of liquidity provision in large tick LOBs. Its implications are diverse. Our generalization of MRR implies that the traders' consensus defines a fundamental price on scales below the tick size. Changes in the fundamental price are reflected in changes of the volumes in the LOB, which allows us to define a proxy of the fundamental price based on the liquidity at the best. This is in stark contrast to the zero intelligence approach embraced by a different branch of the literature on the subject \citep{Farmer:2005predictive, ContStoikov}.

Several empirical tests on high frequency data support our model's core results (in particular Eq.~(\ref{eq:xt3})), in that the predicted relationship between price impact and trade sign correlations holds accurately far beyond the LOB's price resolution. Whereas the core of our framework is very successful, we are also aware that the performance of the proxy $\hat p_t$ defined in Sec.~\ref{sec:proxy} is acceptable but not overwhelming. While the dynamics of $\hat p_t$ are roughly in line with what we expect from the fundamental price, i.e. Eq.~(\ref{eq:pt}), $\hat p_t$ incorporates $\approx 80\%$ of the information contained in the volume imbalance at the best; this is encouraging but not outstanding. 

Other aspects of our paper call for further research, as well. Whereas we are able to link the dynamics of the volume imbalance to the location of the approximate fundamental price, our work does not develop a criteria to determine the \emph{absolute} level of liquidity in the LOB. The relationship between overall liquidity and the tick size remains an important unsolved problem in applied market microstructure.
Second, this paper has ignored MRR's second equation~(\ref{eq:st}) which relates the spread to the response function according to $s = 2 R(1)/(1 - C(1))$. Whereas the implied spread overestimates the realized spread a little, we do not observe a significant difference between small and large tick stocks. This is again very surprising and suggests that the permanent impact of a market order, $R(\infty) = G$, is on large tick LOBs approximately equal to $\tau/2$, irrespective of other properties of the stock, such as its price or volatility. Do liquidity takers only submit market orders if the value of their private information exceeds the trading costs? Why do order flow correlations and the response function conspire in such a way that Eq.~(\ref{eq:st}) is approximately satisfied, even on large tick stocks? We must leave these questions unanswered here. 

This paper postulates the existence of a fundamental price which is by construction totally unpredictable. Is this consensus price generated by a crowd of equally rational agents? This does not seem the case. By zooming into the instant when a queue depletes and the price changes, one finds that in some cases (approximately $25\%$) the depleted queue is immediately refilled: Liquidity providers do not always agree whether a mid price change is acceptable, and the diverging opinions create significant fluctuations. How this variety of perceptions merge eventually into a consensus we cannot tell; we do not derive our model from first principles. But as a \emph{phenomenological model} it can improve our understanding of financial markets.

\section*{Acknowledgements}
We thank Jonathan Donier, Jean-Philippe Bouchaud and Charles-Albert Lehalle for their many important comments and insights. We acknowledge discussions with Mathieu Rosenbaum, Martin Gould, Rama Cont and Arseniy Kukanov.

\begin{landscape}

\begin{table}
  \centering
  \begin{tiny}
  \begin{tabular}{c|c|c|c}
    
    ticker & mean price & mean spread & trade sign  \\
           &  $[\$]$    &      $[\$]$ &        correlation $\Ex\epsilon_t\epsilon_{t+1}$\\
    \hline

    SIRI &  3.87  &0.0108 & 0.89 \\
    AMAT & 19.11  &0.0116 & 0.62 \\
      MU & 20.71  &0.0124 & 0.52 \\
    SYMC & 22.57  &0.0115 & 0.54 \\
    NVDA & 23.04  &0.0119 & 0.53 \\
     MAT & 24.98  &0.0120 & 0.50 \\ 
    ATVI & 26.47  &0.0121 & 0.50 \\
    GE   & 26.60  &0.0123 & 0.62 \\
    CSCO & 27.81  &0.0113 & 0.69 \\
      JD & 29.34  &0.0147 & 0.38 \\
      CA & 29.72  &0.0120 & 0.45 \\
   DISCA & 30.51  &0.0127 & 0.41 \\
     CSX & 31.08  &0.0123 & 0.45 \\ 
    FOX & 31.21  &0.0117 & 0.47 \\
    INTC & 31.96  &0.0116 & 0.63 \\
     VOD & 34.55  &0.0117 & 0.38 \\
    AMTD & 35.48  &0.0142 & 0.38 \\
    TMUS & 35.96  &0.0165 & 0.39 \\
    IBKR & 37.11  &0.0421 & 0.43 \\ 
    LMCA & 37.94  &0.0153 & 0.35 \\
   YHOO  &38.20  &0.0122  &0.47 \\
    EBAY & 38.33 & 0.0121 & 0.48 \\
    WFM  &39.34  &0.0133  &0.38 \\
    MDLZ & 40.48 & 0.0119 & 0.49 \\ 
    ORCL & 40.64 & 0.0119 & 0.48 \\
    FAST & 40.90 & 0.0140 & 0.37 \\
   XLNX  &43.63  &0.0130  &0.35 \\
    LLTC & 44.34 & 0.0135 & 0.37 \\
      AAL&  44.74&  0.0156&  0.40 \\
    MSFT & 46.26 & 0.0118 & 0.57 \\
    PAYX & 48.91 & 0.0142 & 0.36 \\
    NDAQ & 51.66 & 0.0213 & 0.38 \\
     TXN & 53.67 & 0.0124 & 0.39 \\
    CINF & 53.97 & 0.0240 & 0.30 \\
   NCLH  &54.32  &0.0326  &0.43 \\
     ADSK&  55.53&  0.0201&  0.33 \\
    VIAB & 55.93 & 0.0175 & 0.36 \\
     MYL & 56.10 & 0.0232 & 0.42 \\
     KLAC  & 58.64  &0.0234&  0.37 \\
     CMCSA &  58.90 & 0.0120&  0.46 \\
     ADI   &59.35  &0.0194 & 0.36 \\
     PCAR  & 59.45 & 0.0177&  0.34 \\
     QCOM  & 61.89 & 0.0125&  0.40 \\
     SBUX  & 61.93 & 0.0151&  0.38 \\
     CTSH  & 62.05 & 0.0166&  0.36 \\
     CTRP  & 62.50 & 0.0687&  0.60 \\
     JPM   &63.58  &0.0124 & 0.38 \\ 
     ROST  & 64.88 & 0.0281&  0.41 \\
     CERN  & 66.15 & 0.0259&  0.38 \\
     DISH  & 66.57 & 0.0421&  0.45 \\

  \end{tabular}
  \quad
  \begin{tabular}{c|c|c|c}

    ticker & mean price & mean spread & trade sign  \\
    &  $[\$]$    &      $[\$]$ &        correlation $\Ex\epsilon_t\epsilon_{t+1}$\\
    \hline

     SNDK  & 67.33 & 0.0287&  0.44 \\
     CHRW  & 68.09 & 0.0291&  0.46 \\
     CTXS  & 68.75 & 0.0304&  0.42 \\
     NTRS  & 71.82 & 0.0292&  0.39 \\
     DLTR  & 74.40 & 0.0257&  0.41 \\
     MAR   &75.55  &0.0275 & 0.40 \\ 
     LRCX  & 76.06 & 0.0402&  0.51 \\
     TROW  & 77.30 & 0.0280&  0.38 \\
     ADBE  & 80.33 & 0.0243&  0.37 \\ 
     CHKP  & 81.92 & 0.0405&  0.52 \\
     XOM   &82.49  &0.0139 & 0.36 \\
     WDC   &83.63  &0.0428 & 0.50 \\
     FISV  & 83.70 & 0.0323&  0.39 \\
     ADP   &84.22  &0.0230 & 0.36 \\
     WBA   &85.01  &0.0300 & 0.41 \\
     ESRX  & 85.93 & 0.0232&  0.41 \\
     TSCO  & 87.30 & 0.0679&  0.53 \\
     FB   &87.53  &0.0147  &0.37 \\
     SWKS &  88.38&  0.0543&  0.55 \\
     NXPI &  90.60&  0.0558&  0.62 \\
     CME  & 93.65 & 0.0428 & 0.52 \\
     INTU &  96.73&  0.0422&  0.48 \\ 
     INCY & 100.40&  0.1877&  0.70 \\
     GILD & 106.16&  0.0258&  0.45 \\
     EXPE & 106.48&  0.0746&  0.59 \\
     BMRN & 115.06&  0.201 & 0.72 \\
     CELG & 118.02&  0.0549&  0.59 \\
     AAPL & 119.72&  0.0138&  0.33 \\
     VRTX & 123.14&  0.166 & 0.71 \\
     AVGO & 124.10&  0.085 & 0.64 \\
     NTES & 127.10&  0.295 & 0.71 \\
     MNST & 135.37&  0.118 & 0.60 \\ 
     HSIC & 143.01&  0.134 & 0.59 \\
     COST & 147.31&  0.052 & 0.56 \\
     ULTA & 156.02&  0.178 & 0.71 \\
     AMGN & 157.73&  0.080 & 0.59 \\
     ALXN & 176.51&  0.239 & 0.71 \\
     CHTR  & 179.60&  0.198 & 0.70 \\
     BIDU  & 187.66&  0.144 & 0.67 \\
     NFLX  & 188.93&  0.256 & 0.64 \\
     ILMN  & 189.26&  0.254 & 0.73 \\
     TSLA  & 227.40&  0.214 & 0.67 \\
     ORLY  & 229.37&  0.239 & 0.65 \\
     EQIX  & 259.21&  0.333 & 0.73 \\
     BIIB  & 344.48&  0.394 & 0.74 \\ 
     AMZN  & 452.08&  0.326 & 0.72 \\
     REGN  & 491.74&  0.975 & 0.76 \\
     ISRG  & 505.92&  0.989 & 0.73 \\
     GOOG  & 591.23&  0.392 & 0.75 \\
     PCLN  &1211.77&  1.567 & 0.77 
     
  \end{tabular}
  \end{tiny}
  \caption{Summary statistics of the pool of $100$ small, medium and large tick stocks traded on Nasdaq during 2015. The stocks are ordered with ascending price. Averages are calculated in transaction time by using the prevailing quotes at market order arrivals.}
  \label{tab}
\end{table}
\end{landscape}


\bibliographystyle{abbrvnat}
\bibliography{bibli}

\end{document}